\documentclass[11pt]{article}
\pdfoutput=1

\usepackage{tikz}
\usepackage{euscript}
\usepackage{amssymb}
\usepackage{amsfonts}
\usepackage{amsbsy}
\usepackage{epsfig}
\usepackage{amsthm}
\usepackage{amscd}
\usepackage{amstext}
\usepackage{verbatim}
\usepackage{amsmath}
\usepackage{cancel}
\usepackage{capt-of}
\usepackage{empheq}
\usepackage{subfigure}

\usepackage[nosort]{cite}
\usepackage{bm}
\usepackage{authblk}
\usepackage{esint}
\usepackage[T1]{fontenc}
\usepackage{mathdots}
\usepackage[centertableaux]{ytableau}

\usepackage[utf8]{inputenc}
\usepackage{graphicx}
\usepackage{physics}
\usepackage{mathtools}
\usepackage{bbold}

\usepackage{setspace}
\usepackage{colortbl}
\usepackage{xcolor}

\usepackage[pdftex,linktoc=all]{hyperref}
\hypersetup{ 
colorlinks=true, 
linkcolor=black, 
citecolor=red, 
}

\def\ben{\begin{equation}}
\def\een{\end{equation}}
\def\half{{\textstyle{\frac{1}{2}}}}

   \let\x=\xi 
 \let\t=\tau

\let\w=\omega

\def\nn{\nonumber}

\let\pa=\partial
\def\be{\begin{equation}}
\def\ee{\end{equation}}
\def\beq{\begin{equation}}
\def\eeq{\end{equation}}
\def\ba{\begin{array}}
\def\ea{\end{array}}

\def\dalemb#1#2{{\vbox{\hrule height .#2pt
       \hbox{\vrule width.#2pt height#1pt \kern#1pt
               \vrule width.#2pt}
       \hrule height.#2pt}}}

\newcommand{\bea}{\begin{eqnarray}}
\newcommand{\eea}{\end{eqnarray}}

\def\vep{{\varepsilon}}

\makeatletter
\newcommand*\bigcdot{\mathpalette\bigcdot@{.5}}
\newcommand*\bigcdot@[2]{\mathbin{\vcenter{\hbox{\scalebox{#2}{$\m@th#1\bullet$}}}}}
\makeatother

\renewcommand{\eqref}[1]{(\ref{#1})}

\def\Z{{{\mathbb Z}}}

\def\ocal{{\mathcal{O}}}

\makeatletter
\newcommand{\curvedsquare}{%
  \mathchoice
    {\@curvedsquare{0.8em}{0.06em}}% \displaystyle
    {\@curvedsquare{0.8em}{0.06em}}% \textstyle
    {\@curvedsquare{0.6em}{0.05em}}% \scriptstyle
    {\@curvedsquare{0.4em}{0.04em}}% \scriptscriptstyle
}
\newcommand{\@curvedsquare}[2]{%
  \mathord{\vcenter{\hbox{\tikz[x=#1, y=#1]{
    \draw[line width=#2, line join=round]
      (0,0) -- (0,1)                 % Left straight line (up)
      to[bend right=30] (1,1)        % Top curved line (right & down)
      -- (1,0)                       % Right straight line (down)
      to[bend right=30] (0,0);       % Bottom curved line (left & up)
  }}}}%
}
\makeatother

\begin{document}

\begin{center}
\begin{tikzpicture}
   \draw [black,thick,domain=6:6.5] plot ({\x}, {0.85*sqrt{ ((\x)*(\x) - 36)} });
   \draw [black,thick,domain=6:6.5] plot ({\x}, {-0.85*sqrt{ ((\x)*(\x) - 36)} });
   \draw [black,thick,domain=-6.5:-6] plot ({\x}, {0.85*sqrt{ ((\x)*(\x) - 36)} });
   \draw [black,thick,domain=-6.5:-6] plot ({\x}, {-0.85*sqrt{ ((\x)*(\x) - 36)} });
   \draw [black,thick] (-6,-0.05) -- (-6,0.05);
   \draw [black,thick,domain=-6.5:6.5, samples=100, smooth] plot ({\x},
   {2.125*sqrt{ ((35+\x * \x)/(35+42.25)) }});
   \draw [black,thick,domain=-6.5:6.5, samples=100, smooth] plot ({\x},
   {-2.125*sqrt{ ((35+\x * \x)/(35+42.25)) }});
   \node[align=center] at (0,0) {{\huge Holographic Banners} \\
   {\large Matthew~J.~Blacker and Sean~A.~Hartnoll} \\
{\large \it DAMTP, University of Cambridge, Cambridge CB3 0WA, UK}};
\end{tikzpicture}
\end{center}

\vspace{2cm}
\begin{abstract}

This paper is concerned with eternal AdS black holes.  The quantum cosmological future and past interior states of the black hole may be placed on an equal footing to the left and right AdS boundary data by considering the on-shell bulk action as a function of the left/right/future/past data: $S[\phi^{(0)L},\phi^{(0)R},\phi^{(0)F},\phi^{(0)P}]$. We call this object a holographic banner, and it obeys the Hamilton-Jacobi equation with respect to all four of its arguments. We compute the holographic banner for a scalar field in an AdS black hole background explicitly and use it to construct the semiclassical state in the future interior obtained from a thermofield double state in the past evolved by arbitrary time- and space-dependent boundary sources. When the spacetime itself is dynamical we explain how the holographic banner gives, in principle, a map from boundary data to near-singularity semiclassical quantum cosmology following chaotic BKL dynamics. We obtain the timescale for the BKL dynamics to ergodically mix the future interior quantum state, given a quantum variance in the past state or a classical ensemble of boundary theories.

\end{abstract}

\newpage

\tableofcontents

\newpage

\section{Introduction}

The best understood instances of holographic duality start with a quantum mechanical theory and dynamically build an emergent `radial' spatial dimension. The quantum mechanical `boundary' time, however, does not extend past any horizons in the emergent spacetime.
It is therefore challenging to access important physics beyond the horizon, such as black hole singularities or the future of de Sitter spacetime.
Previous works have shown that the singularity of certain black holes can be probed by analytic continuation in the boundary time and/or energy \cite{Fidkowski:2003nf, Festuccia:2005pi, Frenkel:2020ysx, Grinberg:2020fdj, Afkhami-Jeddi:2025wra, Dodelson:2025jff}. This approach
avoids grappling with the holographic emergence of time in the black hole interior, where the radial direction becomes timelike. In generic situations, with fewer symmetries, the emergent interior time should be logically distinct from the pre-existing boundary time. Recent approaches to this dichotomy include \cite{deBoer:2022zps, Leutheusser:2025zvp}.

It was explained in \cite{Hartnoll:2022snh} that the exterior and interior could be treated uniformly within a Hamilton-Jacobi description of the classical dynamics.
By using a relational `clock' on radial slices, such as the redshift or the volume of the spatial sphere, the dynamics develops smoothly from the exterior boundary, through the horizon and to the interior singularity. The quantum version of this relational description is the Wheeler-DeWitt equation \cite{DeWitt:1967yk}.
On an interior slice this equation determines the wavefunction of the black hole interior. On an exterior slice it determines the holographic renormalisation of the dual CFT partition function \cite{deBoer:1999tgo, deHaro:2000vlm, Heemskerk:2010hk, Faulkner:2010jy, McGough:2016lol, Hartman:2018tkw}. The framework of \cite{Hartnoll:2022snh} has been extended to charged black holes \cite{Blacker:2023ezy} and to de Sitter horizons \cite{Blacker:2023oan}. All of these works, however, were limited to a time-independent `mini-superspace' and did not incorporate the two-sidedness of black hole spacetimes. In the present paper we introduce mathematical objects --- holographic banners --- that combine the interior and exterior in full generality.

\begin{figure}[h]
\centering
\includegraphics[width=0.65\textwidth]{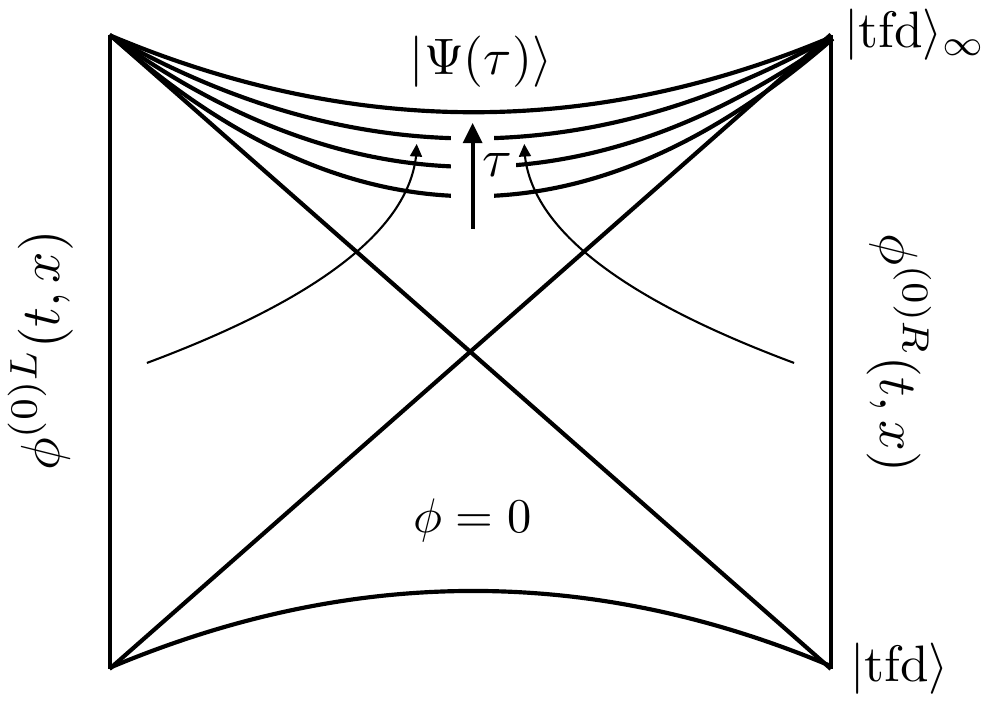}
\caption{A {\bf holographic banner}: Boundary couplings $\phi^{(0) L/R}$ evolve the boundary thermofield double state $|\text{tfd}\rangle$ to a late-time boundary state $|\text{tfd}\rangle_\infty$. The boundary couplings source a bulk field $\phi$ which is classically zero in the past interior. In the future interior the bulk field is nonzero, due to the sources, and in a semiclassical quantum state $|\Psi\rangle$ on an interior slice. Near the singularity this state evolves in a simple way under a relational interior time $\tau$.}
\label{fig:bigpicture}
\end{figure}

Imagine a canvas banner stretched between two poles. The supporting poles are two AdS boundaries entangled in a thermofield double state, the canvas is the bulk spacetime and the top and bottom of the banner are slices through the bulk. We will be interested in situations where the bulk contains a black hole and the corners of the banner are at $\pm \infty$ boundary time. This means that the bulk slices are in the future and past interior of the black hole. See Fig.~\ref{fig:bigpicture}. At a classical level the natural data associated to this setup is the bulk action as a function of the field values on the four (left/right/future/past) boundaries: $S[\phi^{(0)L},\phi^{(0)R},\phi^{(0)F},\phi^{(0)P}]$. This quantity obeys the Hamilton-Jacobi equation (including modifications due to boundary counterterms) with respect to each of its arguments independently, and has been considered previously in \cite{Skenderis:2008dh, Skenderis:2008dg}. We will call it a holographic banner.

The quantum mechanical bulk state evolves from the past slice to the future slice subject to the boundary sources $\phi^{(0) L/R}$ \cite{Balasubramanian:1998sn}. These sources can depend on both boundary time and space. Semiclassical bulk wavepackets are obtained by integrating $e^{i S[\phi^{(0)L},\phi^{(0)R},\phi^{(0)F},\phi^{(0)P}]}$ against a distribution of $\phi^{(0)P}$ or $\phi^{(0)F}$. The boundary sources, in contrast, are kept fixed if there is a fixed boundary theory. It is also natural to consider ensembles of boundary theories in this setup, and we will do so in \S\ref{sec:ensemble} below.

The simplest initial state for a two-sided black hole spacetime is the thermofield double state $|\text{tfd}\rangle$ of the boundary theory \cite{Maldacena:2001kr}. In this state a scalar field $\phi$ in the bulk vanishes classically in the past interior. Corresponding semiclassical states can be built as Gaussian wavepackets that are strongly peaked on $\phi = 0$ throughout the past interior. The classical field is nonzero in the future interior, however, due to the boundary sources. The semiclassical state $|\Psi\rangle$ on the future slice will then be strongly supported on the future classical field, together with a quantum mechanical phase given by evaluating the action on the classical bulk solution. It is important to distinguish $|\Psi\rangle$ from the boundary future state $|\text{tfd}\rangle_\infty$, obtained by time-evolving $|\text{tfd}\rangle$ using the boundary Hamiltonian in the presence of the sources $\phi^{(0) L/R}$. For given fixed sources, $|\text{tfd}\rangle_\infty$ is a unique state. In contrast, $|\Psi\rangle$ evolves relationally as a function of the interior slice, which we label by $\tau$. The setup we have just described is illustrated in Fig.~\ref{fig:bigpicture}. We can also consider more general states in the past, which correspond to deformations of the thermofield double state, cf.~\cite{Balasubramanian:2022gmo}.

As we have stressed, interior evolution with $\tau$ is unrelated to the boundary time evolution by $t$. Interior time evolution with fixed boundaries for the slice is `pure gauge' and governed by the Hamiltonian constraint once gravity is dynamical. This is equivalent to the statement that the interior evolution is relational. That is, $|\Psi(\tau)\rangle$ is the state of the bulk conditioned upon a value $\tau$ for a certain geometric observable that fixes the slice. There are many possible choices for the observable, see e.g.~\cite{Hartnoll:2022snh}.
For a scalar field in a fixed black hole background we will see that there is an especially nice choice such that $\tau \to \infty$ as the slice approaches the future singularity.
For each wavevector mode $\phi_k$ of the scalar field the wavepackets on the slice become
\be\label{eq:state}
\Psi(\phi_k, \tau) = \psi_{p_k,\tau}(\phi_k) \, e^{-i |p_k|^2 \tau} \,,
\ee
where $\psi_{p_k,\tau}$ is strongly peaked on the classical solution $\langle\phi_k\rangle_\psi = p_k \tau$ and $\langle - i \pa_{\phi_k}\rangle_\psi = \bar p_k$. That is, the state (\ref{eq:state}) simply describes a free particle with complex `momentum' $p_k$. This momentum is a function of the boundary data $\phi^{(0) L/R}$ and therefore the holographic banner gives a map
\be\label{eq:map}
\phi^{(0) L/R} \qquad \mapsto \qquad |\Psi(\tau)\rangle \in \bigotimes_k {\mathcal H}_k \,,
\ee
from classical boundary data to interior single-particle states living in the free particle Hilbert spaces ${\mathcal H}_k$. The first technical result of this paper is an explicit form of the map (\ref{eq:map}) for a scalar field in a planar AdS black hole background in general dimension, without taking a near-singularity limit. The holographic banner $S[\phi^{(0)L},\phi^{(0)R},\phi^{(0)F},\phi^{(0)P}]$ is obtained in (\ref{eq:os4}) and (\ref{eq:osfinal}), 
the Gaussian wavepacket is given in (\ref{eq:psifinal}). The quantum mechanical phase, in particular, is obtained in (\ref{eq:scl}).

Our result (\ref{eq:osfinal}) for the holographic banner is given in terms of four independent quantities. The first is the exterior (momentum-space) Green's function for the dual operator, which is a familiar object \cite{Hartnoll:2016apf} giving the relative importance of the normalisable and non-normalisable modes near the AdS boundary. The second is the newly defined interior Green's function, which determines the relative importance of a mode that grows logarithmically towards the singularity and a mode that doesn't. The remaining two quantities, defined in (\ref{eq:DX}), are a phase and magnitude that depend up the relative growth of the modes in the interior and exterior. All of these quantities can be computed explicitly for a scalar field in a three-dimensional BTZ background, and the map (\ref{eq:map}) in that case is given by (\ref{eq:PI}).

\begin{figure}[h]
\centering
\includegraphics[width=0.5\textwidth]{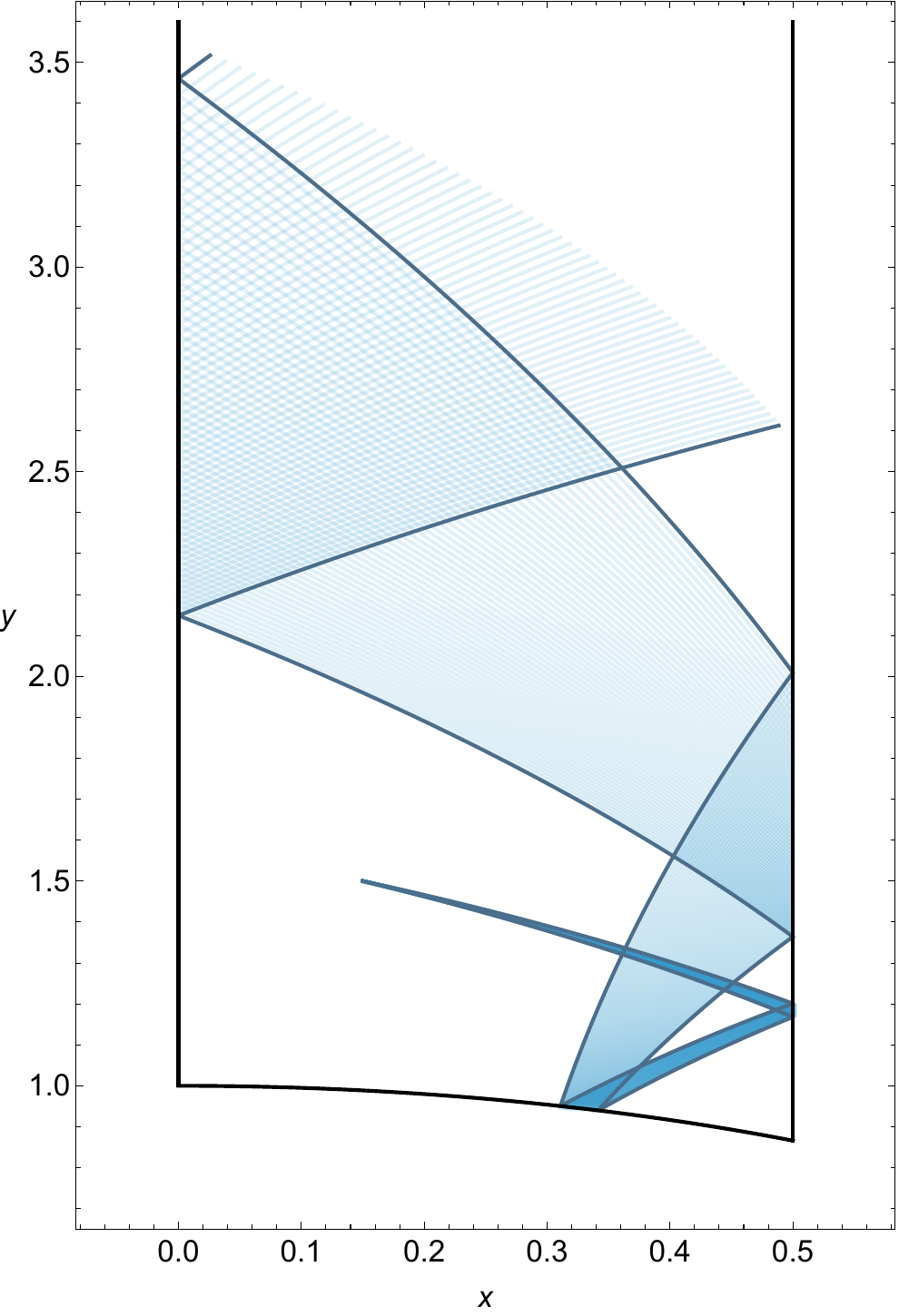}
\caption{The {\bf universe near a BKL singularity} can be mapped, at each point in space separately, to the trajectory of a particle moving within half of the fundamental domain of $SL(2,\Z)$ in $\mathbb{H}_2$. The exponential divergence of nearby geodesics causes a semiclassical wavepacket to spread across the entire domain over the mixing time $\tau_\text{mix}$. The figure illustrates the spreading of a collection of geodesics.}
\label{fig:domain}
\end{figure}

Once backreaction on the metric and gravitational  nonlinearities are incorporated, the near-singularity bulk state will no longer be a product over wavevector modes. Remarkably, however, the BKL scenario \cite{Belinsky:1970ew, Damour:2002et, belinski_henneaux_2017} suggests that there is a large class of near-singularity dynamics where non-interacting single-particle states again emerge. In the BKL scenario distinct spatial points $x$ on the slice decouple towards the singularity. The holographic banner (with a given state in the past) thus now defines a semiclassical map
\be\label{eq:mapbkl}
\phi^{(0)L/R} \qquad \mapsto \qquad |\Psi(\tau)\rangle \in \bigotimes_{x} {\mathcal H}_{x} \,.
\ee
The ${\mathcal H}_x$ Hilbert spaces again describe single-particle states, except that now the particles evolve in $\tau$ following a chaotic hyperbolic billiard dynamics. We give a brief summary of this dynamics and further references in \S\ref{sec:bkl}. A new physical effect that arises is the exponential spreading of the classical hyperbolic trajectories. Combined with the finite volume of the billiard domain, this spreading leads to a mixing time $\tau_\text{mix}$, given in (\ref{eq:mix}), at which the wavepacket becomes uniform across the domain. This mixing is illustrated in Fig.~\ref{fig:domain}.

In \S\ref{sec:bkl} we show that if the initial variance of the wavepacket is set by the Planck scale, then the spreading is not fast enough to mix the wavepacket before the interior universe collapses to Planck size. That is, quantum gravity (or stringy) effects will become important before the semiclassical spreading is complete. However, classical bulk states can be constructed with a quantum variance that is large enough to produce a quantum mechanically smeared interior before the universe reaches the Planck scale. An even larger variance can be achieved by considering an ensemble of boundary theories. This amounts to averaging the boundary sources $\phi^{(0)L/R}$ over some distribution. This distribution can have a classical width, leading to a faster mixing time. Thus the interior states dual to ensembles of boundary theories are expected to be especially universal, captured by ergodic properties of the billiard dynamics.

Irrespectively of the mixing time, recent work has argued for an interior duality --- perhaps in the spirit of a dS/CFT correspondence --- of BKL gravitational dynamics with `primon gas' partition functions \cite{Hartnoll:2025hly, DeClerck:2025mem}. The holographic banner map (\ref{eq:mapbkl}) is then a bridge between the exterior AdS/CFT duality and the interior primon gas duality. To truly make use of this bridge we need to understand whether and how the relational interior time $\tau$ is encoded in the boundary theory. One set of ideas in this direction involve so-called $T^2$ deformations \cite{Belin:2020oib, Araujo-Regado:2022gvw}. A fascinating prospect here is that the boundary theory could give a new perspective on various number-theoretic properties of the primon gas.

In the final discussion \S\ref{sec:dis} we note some future directions including a fully quantum holographic banner defined from a four-sided path integral, a generalisation of the holographic banner to de Sitter spacetime and the possibility of using matrix degrees of freedom to define relational observables in the boundary theory.

\section{Scalar field in an AdS black hole background}

We will obtain an explicit expression for the holographic banner of a scalar field in a $d+2$ dimensional planar AdS black hole background,
\be\label{eq:met}
ds^2 = \frac{L^2}{z^2} \left(-f(z) dt^2 +  \frac{dz^2}{f(z)} + dx_d^2 \right)\,.
\ee
Here $L$ is the AdS radius. While the precise form of $f(z)$ is not too important, for concreteness we can take the planar AdS-Schwarzschild expression
\be\label{eq:f}
f(z) = 1 - \left(\frac{z}{z_\mathcal{H}} \right)^{d+1} \,.
\ee
The horizon is at $z = z_\mathcal{H}$ and the corresponding black hole temperature is
\be\label{eq:temp}
T = \frac{d+1}{4 \pi z_\mathcal{H}} \,.
\ee
The AdS boundary is at $z \to 0$ and the singularity as at $z \to \infty$. For more details about these backgrounds see e.g.~\cite{Hartnoll:2016apf}.

Consider a complex scalar field $\Phi$ in the geometry (\ref{eq:met}), with action given by
\be\label{eq:actPhi}
S = - \int dt dz d^{d}x \sqrt{-g} \Big(\nabla_\mu \bar \Phi \nabla^\mu \Phi + m^2 |\Phi|^2 \Big) \,.
\ee
A Fourier decomposition of the scalar field along the AdS boundary directions is
\be
\Phi = \int \frac{d\omega d^{d}k}{(2\pi)^{d+1}} e^{- i \omega t + i k \cdot x} \phi_{\omega k}(z) \,. \label{eq:fourier}
\ee
Here the integral is over $\omega \in (-\infty,\infty)$.
The action (\ref{eq:actPhi}) then becomes
\be\label{eq:actwk}
S = - L^{d} \int \frac{d\omega d^{d}k}{(2\pi)^{d+1}} \frac{dz}{z^{d}} \left(\left[-\frac{\omega^2}{f} + k^2 + \frac{(L m)^2}{z^2}\right] |\phi_{\omega k}|^2 + f \left|\frac{d\phi_{\omega k}}{dz}\right|^2 \right) \,.
\ee
The equation of motion following from (\ref{eq:actwk}) is
\be\label{eq:eom}
- z^d \frac{d}{dz} \left( \frac{f}{z^d} \frac{d \phi_{\omega k}}{dz} \right) + \left[-\frac{\omega^2}{f} + k^2 + \frac{(L m)^2}{z^2}\right] \phi_{\omega k} = 0 \,.
\ee
This kind of equation has been solved extensively in black hole exteriors. With infalling boundary conditions on the future horizon, the solutions describe relaxation to thermal equilibrium in the boundary theory \cite{Hartnoll:2016apf}. To obtain the holographic banner we will need to solve the equation in both the interior and exterior and match correctly across horizons.

\subsection{Global coordinates and matching across horizons}
\label{sec:hor}

\begin{figure}[h]
\centering
\includegraphics[width=0.48\textwidth]{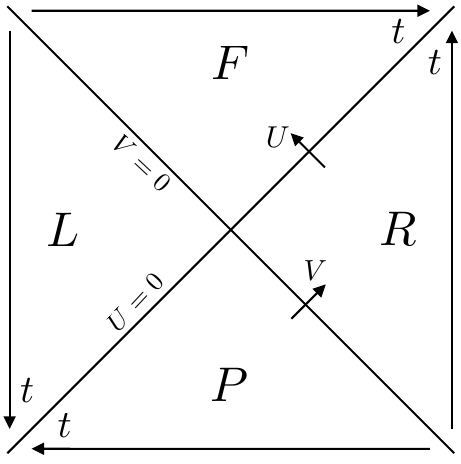}
\caption{Four regions, each covered by separate $z,t$ coordinates. The direction of increasing Schwarzschild $t$ coordinate in each region is shown. The diagonal lines are the horizons, determined by either $U$ or $V$ vanishing. The direction of increasing $U$ and $V$ is also shown.}
\label{fig:regions}
\end{figure}

As is well-known, the black hole horizons split the spacetime up into four regions, labelled by $X = \{R,L,F,P\}$ in Fig.~\ref{fig:regions}. The 
coordinate system (\ref{eq:met}) can be used in any one of these four regions, but the $z,t$ coordinates are distinct in each region. It will be important to review how the different coordinates are related. There is a large literature on matching modes across a horizon, following the classic work by Unruh \cite{Unruh:1976db}. Our task will be easier for the most part --- we wish to determine the continuity conditions relating classical solutions in the different regions.

Define the usual tortoise coordinate in the interior and exterior, respectively, as
\be
z_\star = \int_z^\infty \frac{dw}{f(w)} \,, \qquad z_\star = - \int_0^z \frac{dw}{f(w)}  \,.
\ee
In both cases $z_\star$ runs from $-\infty$ to $0$. In particular, near a horizon $z  \to z_{\mathcal H}$
\be\label{eq:fhor}
f \approx 4 \pi T (z_\mathcal{H} - z) \quad \Rightarrow \quad z_\star \approx \frac{1}{4 \pi T} \log |z_{\mathcal H} - z| \,,
\ee
so that $z_\star \to - \infty$.
Global coordinates $U,V$ are then related to $z,t$ in the four different regions by
\be\label{eq:change}
R:
\left\{\begin{array}{l}
U = - e^{- t + z_\star}\\
V = e^{ t + z_\star}
\end{array} \right. \;\;
L:
\left\{\begin{array}{l}
U = e^{- t + z_\star} \\
V = - e^{ t + z_\star}
\end{array} \right. \;\;
F:
\left\{\begin{array}{l}
U = e^{- t + z_\star}\\
V = e^{ t + z_\star}
\end{array} \right. \;\;
P:
\left\{\begin{array}{l}
U = - e^{- t + z_\star}\\
V = - e^{ t + z_\star}
\end{array} \right. \,.
\ee
The horizons are at $U=0$ and $V=0$, so that $UV$ changes sign across each horizon. The metric (\ref{eq:met}) in global coordinates is
\be\label{eq:global}
ds^2 = \frac{L^2}{z^2} \left(\frac{f \, dUdV}{UV} + dx_d^2\right) \,.
\ee
There is no coordinate singularity in (\ref{eq:global}) at the horizons, as $UV$ vanishes together with $f$. Note that $f/(UV)$ is negative everywhere.

The full action is a sum of the action in the four regions, each given by (\ref{eq:actPhi}) or (\ref{eq:actwk}), with no relative minus signs. To avoid potential confusions, 
we are {\it not} going to be computing Schwinger-Keldysh correlation functions,
which require a contour in complex time that does involve a change in orientation and a corresponding minus sign \cite{Herzog:2002pc, Skenderis:2008dh, Skenderis:2008dg, Liu:2018crr}. Instead, we are interested in preparing a bulk state. This requires the on-shell action of the global bulk Lorentzian spacetime, as we see in more detail below, in which the action of the different regions add (cf.~\cite{Brown:2015lvg}). For convenience we will stick with the Schwarzschild $t$ coordinate throughout instead of introducing separate left and right times. Recall that the $t$ coordinate runs in different directions between the left and right and between the future and past, as shown in Fig.~\ref{fig:regions}. Note that redefining the time coordinate would not introduce any minus signs in the relative contribution of the different regions.

To match scalar field solutions between the various regions, consider the near-horizon behaviour of the field. In any region $X$ the modes behave near the horizon $z \to z_\mathcal{H}$ as
\be\label{eq:nh}
\phi^X_{\omega k} \approx A^X_{\omega k} |z_\mathcal{H} - z|^{ - i \omega/(4 \pi T)} + B^X_{\omega k} |z_\mathcal{H} - z|^{i \omega/(4 \pi T)} \,.
\ee
Here $A^X_{\omega k}$ and $B^X_{\omega k}$ are coefficients.
The behaviour (\ref{eq:nh}) follows from the near-horizon expansion (\ref{eq:fhor}) and solving the equations of motion.
Reinstating the time dependence,
\be\label{eq:phiUV}
\phi^X_{\omega k} e^{- i \omega t} \approx A^X_{\omega k} |V|^{-i \omega} + B^X_{\omega k} |U|^{i \omega} \,.
\ee
Where we used the fact that the $\pm$ signs in the change of variables (\ref{eq:change}) determine the sign of $U$ and $V$.

Consider first the $U=0$ horizon separating the $F/R$ and $L/P$ regions. This horizon is parametrised by the $V$ coordinate. For continuity of the wave profile across the horizon we must therefore match the $V$ dependence in (\ref{eq:phiUV}), so that
\be\label{eq:A1}
A^F_{\omega k} = A^R_{\omega k} \,, \qquad A^P_{\omega k} = A^L_{\omega k} \,.
\ee
Modes associated to a nonzero $B_{\omega k}$ are singular on the $U=0$ horizon, as they oscillate infinitely many times upon approach. These are `outgoing' modes that are causally disconnected on the two sides of the horizon, specified by independent initial data. They are therefore not subject to a continuity relation like (\ref{eq:A1}). Regular fields will be obtained from suitable wavepackets of the individual modes, after integration over $\omega$. We give an explicit example of this phenomenon, as well as of the matching conditions, in Appendix \ref{app:match}.

Applying the same logic to the $V=0$ horizon we obtain the further matching conditions
\be\label{eq:A2}
B^P_{\omega k} = B^R_{\omega k} \,, \qquad B^L_{\omega k} = B^F_{\omega k} \,.
\ee
These must be imposed together with (\ref{eq:A1}).

\subsection{Near-boundary and near-singularity behaviour}
\label{sec:bdy}

In the previous section we introduced coefficients $A_{\omega k}^X$ and $B_{\omega k}^X$ that characterise the near-horizon behaviour of the fields in each region. The next step is to relate these coefficients 
to boundary data at the left and right AdS boundaries and the future and past singularities. This is necessary in order to evaluate the action as a function of boundary data. In each region, the near-horizon solutions (\ref{eq:nh}) can be extended away from the horizon as
\be\label{eq:full}
\phi^X_{\omega k} = A^X_{\omega k} \phi_{\omega k}^{X-}(z) + B^X_{\omega k} \phi_{\omega k}^{X+}(z)\,.
\ee
Here $\phi_{\omega k}^{X\pm}(z) \approx |z_\mathcal{H} - z|^{ \pm i \omega/(4 \pi T)}$ near the horizon and are solutions to the equation of motion (\ref{eq:eom}) for all $z$.

Consider first $X = \{L,R\}$. The near-boundary $z \to 0$ behaviour is the familiar one from AdS/CFT \cite{Gubser:1998bc, Witten:1998qj}
\be\label{eq:nb}
\phi_{\omega k}^{X\pm}(z) \approx \frac{N_{\omega k}^{X\pm}}{L^{d/2}}\left(z^{d+1-\Delta} + \frac{G_{\omega k}^{X\pm}}{2 \Delta - (d+1)} z^{\Delta} \right) \,.
\ee
As usual $\Delta (\Delta - (d+1)) = (m L)^2$. We will consider so-called standard quantisation, in which $\Delta > d+1-\Delta$ and hence the first term in (\ref{eq:nb}) dominates near the boundary. The coefficient $N_{\omega k}^{X\pm}$ is then the boundary value of the mode, in the AdS/CFT sense, while $G_{\omega k}^{X\pm}$ are the retarded and advanced boundary Green's functions \cite{Son:2002sd, Hartnoll:2016apf}.

Now let $X = \{P,F\}$. The near-singularity $z \to \infty$ behaviour has the following form for $d>1$ (and for $d=1$ with $k=0$),
\be\label{eq:ns}
\phi_{\omega k}^{X\pm}(z) \approx z_{\mathcal H}^{(d+1)/2} \frac{N_{\omega k}^{X\pm}}{L^{d/2}}\left(1 + G_{\omega k}^{X\pm}\log \frac{z}{z_X} \right) \,.
\ee
Here we introduced the coefficients $N_{\omega k}^{X\pm}$ and $G_{\omega k}^{X\pm}$, in analogy to (\ref{eq:nb}). The boundary is at $z = z_X$, with $z_X$ large but finite, so that again $N_{\omega k}^{X\pm}$ are the boundary values of the modes. The logarithmic growth towards the singularity is an indication that the Schwarzschild singularity is unstable once the scalar field backreacts \cite{Frenkel:2020ysx, Doroshkevich:1978aq, Fournodavlos:2018lrk}. The logarithmic running forces us to explicitly introduce the cutoff $z_X$ in (\ref{eq:ns}), unlike for the near-boundary case (\ref{eq:nb}). This can be thought of as an `anomaly' in the holographic banner, analogous to the familiar anomalies in correlation functions that arise in holographic renormalisation \cite{deHaro:2000vlm}.

The wave equation (\ref{eq:eom}) only depends on $\omega^2$ and is real, while the near-horizon boundary conditions, below (\ref{eq:full}), depend on $\pm i \omega$.
It follows that
\be\label{eq:N1}
N_{\omega k}^{X+} = N_{-\omega k}^{X-} = \bar N_{\omega k}^{X-} = \bar N_{-\omega k}^{X+} \,, \qquad G_{\omega k}^{X+} = G_{-\omega k}^{X-} = \bar G_{\omega k}^{X-} = \bar G_{-\omega k}^{X+} \,.
\ee
Furthermore, the right and left coefficients, and also the future and past coefficients, are obtained by solving the same equations with the same initial conditions and hence\footnote{\label{foot:unequal}Because of the logarithmic anomaly in the near-singularity behaviour (\ref{eq:ns}), the future/past equalities in (\ref{eq:N2}) only hold when the cutoffs are taken to be equal: $z_P = z_F$. We will make this choice throughout. If this condition is relaxed one has instead that $N^{P \pm}_{\omega k} = N^{F \pm}_{\omega k} (1 + G^{F \pm}_{\omega k} \log \frac{z_P}{z_F})$ and $G^{P \pm}_{\omega k} = G^{F \pm}_{\omega k}/(1 + G^{F \pm}_{\omega k} \log \frac{z_P}{z_F})$. Many of the expressions below become more cumbersome if these more general relations are used, but the logic remains the same.}
\be\label{eq:N2}
N_{\omega k}^{R\pm} = N_{\omega k}^{L\pm} \,, \qquad N_{\omega k}^{F\pm} = N_{\omega k}^{P\pm} \,, \qquad
G_{\omega k}^{R\pm} = G_{\omega k}^{L\pm} \,, \qquad G_{\omega k}^{F\pm} = G_{\omega k}^{P\pm} \,.
\ee

A further constraint on the asymptotic coefficients is obtained from the presence of a conserved current
\be\label{eq:j}
j \equiv \text{Im} \left(\bar \phi \frac{f}{z^d} \frac{d\phi}{dz}\right) \,, \qquad \frac{dj}{dz} = 0 \,. 
\ee
Conservation follows directly from the wave equation (\ref{eq:eom}) in the way that is familiar from the probability current in  quantum mechanics. Evaluating the current (\ref{eq:j}) at the horizon for the $\phi^{R\pm}$ modes, using (\ref{eq:fhor}), and also at the boundary, using (\ref{eq:nb}), gives
\be
j^{R\pm} = \mp \frac{\omega}{z_{\mathcal H}^d} = \frac{1}{L^d} \left|N_{\omega k}^{R\pm}\right|^2 \text{Im} \, G_{\omega k}^{R\pm} \,. \label{eq:jr}
\ee
And similarly,
\be\label{eq:jf}
j^{F\pm} = \mp \frac{\omega}{z_{\mathcal H}^d} = - \frac{1}{L^d} \left|N_{\omega k}^{F\pm}\right|^2 \text{Im} \, G_{\omega k}^{F\pm} \,.
\ee
Equating the previous two expressions gives
\be\label{eq:rel}
\left|N_{\omega k}^{R-}\right|^2 \text{Im} \, G_{\omega k}^{R-}  = \left|N_{\omega k}^{F+}\right|^2 \text{Im} \, G_{\omega k}^{F+} \,.
\ee
Thus the current at the singularity equals the current near the boundary.

\section{The holographic banner}

\subsection{The on-shell action}

The on-shell action for a configuration that is smooth in the interior of the banner (as noted above, this will involve building a wavepacket out of the individual frequency modes) is a total derivative that can be expressed in terms of boundary data
\be\label{eq:Sos}
S =  - L^d \int \frac{d\omega d^{d}k}{(2\pi)^{d+1}} \left[ \frac{f}{z^d} \bar \phi_{\omega k} \frac{d\phi_{\omega k}}{dz}\right]_{z=z_L,z_R}^{z=z_F,z_P} \,.
\ee
The notation here means that the $z_F$ and $z_P$ terms come with a plus sign while the $z_L$ and $z_R$ terms come with a minus sign. Recall that we will be taking $z_F, z_P \to \infty$ while $z_L,z_R \to 0$.
Accounting for the sign of $f$, then, all four terms in (\ref{eq:Sos}) have the same sign. The on-shell action is necessarily real, which is not manifest in (\ref{eq:Sos}).

In order to relate the $L/R$ contributions in (\ref{eq:Sos}) cleanly to CFT quantities, it is useful to add a local boundary counterterm to the action (\ref{eq:actPhi}), see e.g.~\cite{Hartnoll:2016apf},
\be\label{eq:Sbdy}
S_\text{bdy} = \frac{\Delta - (d+1)}{L} \int d^{d+1}x \sqrt{-\gamma} \, |\Phi|^2 \,.
\ee
This term, which is intrinsic to the boundary, removes local divergences from the on-shell action due to the asymptotically AdS boundaries.

We may now use (\ref{eq:Sos}) to evaluate the on-shell action on the full solutions (\ref{eq:full}). Taking the
near-boundary limits $z \to 0$ (for the $L/R$ regions) or $z \to \infty$ (for the $P/F$ regions), we may then use the near-boundary expansions (\ref{eq:nb}) and (\ref{eq:ns}) to obtain the action as a function of boundary data.
Recalling (\ref{eq:full}), it will be convenient to introduce the full `boundary values' of the fields
\be\label{eq:Ph0}
\phi^{(0)X}_{\omega k} \equiv A^X_{\omega k} N_{\omega k}^{X-} + B^X_{\omega k} N_{\omega k}^{X+} \,.
\ee
The on-shell action can then be written as $S + S_\text{bdy} = \sum_X S^X$, with no minus signs, where the contribution from each region
\be\label{eq:SX}
S^X =  \int \frac{d\omega d^{d}k}{(2\pi)^{d+1}}
\left( \bar \phi_{\omega k}^{(0)X} \left[A_{\omega k}^{X} N_{\omega k}^{X-} G_{\omega k}^{X-} + B_{\omega k}^{X}  N_{\omega k}^{X+} G_{\omega k}^{X+}  \right]\right) \,.
\ee
Near the AdS boundaries we have included the counterterm (\ref{eq:Sbdy}) and used the assumption that $\Delta > d+1-\Delta$ to drop a subleading term. At the boundary near the singularity we have used $\log \frac{z_X}{z_X} = 0$. We will recall in (\ref{eq:pis}) below that the term in square brackets in (\ref{eq:SX}) is nothing but the momentum conjugate to $\bar \phi_{\omega k}^{(0)X}$.

Now we use the relations (\ref{eq:A1}), (\ref{eq:A2}) and (\ref{eq:Ph0}) to simplify the on-shell action (\ref{eq:SX}). Taken together these are eight equations. We solve these to express the eight variables $A^X_{\omega k}$ and $B_{\omega k}^X$ in terms of the $\phi_{\omega k}^{(0)X}$, and thereby eliminate $A^X_{\omega k}$ and $B_{\omega k}^X$ from (\ref{eq:SX}). Thus we obtain, summing over the four regions, the full on-shell action as a function of boundary data:
\be
S\left[\phi_{\omega k}^{(0)L},\phi_{\omega k}^{(0)R},\phi_{\omega k}^{(0)F},\phi_{\omega k}^{(0)P}\right]  =  \int \frac{d\omega d^{d}k}{(2\pi)^{d+1}} s_{\omega k} \,. \label{eq:os4}
\ee
The on-shell action density $s_{\omega k}$ can be written in an elegant way
in terms of the vectors of `external' and `internal' boundary data
\be\label{eq:vv}
v_{\omega k}^{\text{ext}} \equiv \left( \begin{array}{c}
  \phi_{\omega k}^{(0)L} \\
  \phi_{\omega k}^{(0)R}
  \end{array}\right) \,, \qquad 
v_{\omega k}^{\text{int}} \equiv \left( \begin{array}{c}
  \phi_{\omega k}^{(0)F} \\
  \phi_{\omega k}^{(0)P}
  \end{array}\right) \,.
\ee
For a transparent notation we will also introduce the `external' and `internal' Green's functions
\be
G_{\omega k}^\text{ext} \equiv G_{\omega k}^{R-} \,, \qquad G_{\omega k}^\text{int} \equiv G_{\omega k}^{F+} \,.
\ee
We will furthermore package the relative growth of the field in the interior and exterior into
a phase and the ratio
\be\label{eq:DX}
e^{i \Theta_{\omega k}} \equiv \frac{N_{\omega k}^{F+} N_{\omega k}^{R-}}{\left|N_{\omega k}^{F+} N_{\omega k}^{R-} \right|} \,, \qquad X_{\omega k} \equiv \left|\frac{N_{\omega k}^{R-}}{N_{\omega k}^{F+}} \right| \,.
\ee
Finally, we will introduce the matrix of phases
\be
M_{\omega k} \equiv \left(
\begin{array}{cc}
e^{i \Theta_{\omega k}} & e^{- i \Theta_{\omega k}} \\
e^{-i \Theta_{\omega k}} & e^{i \Theta_{\omega k}}
\end{array}
\right) =
\left(
\begin{array}{cc}
-1 & 1 \\
1 & 1
\end{array}
\right)
\left(
\begin{array}{cc}
i \sin{\Theta_{\omega k}} & 0 \\
0 & \cos{\Theta_{\omega k}}
\end{array}
\right)
\left(
\begin{array}{cc}
-1 & 1 \\
1 & 1
\end{array}
\right)\,.
\ee
Using these quantities we have that
\begin{align}
s_{\omega k} \; = \; &
\left|v_{\omega k}^{\text{ext}} \right|^2 \text{Re} \, G_{\omega k}^{\text{ext}} + v_{\omega k}^{\text{ext}\,\dagger} \cdot
 \frac{i M_{\omega k}^2}{\det M_{\omega k}}\cdot v_{\omega k}^{\text{ext}}\,\, \text{Im} \, G_{\omega k}^{\text{ext}} \nonumber  \\
+ \; & \; \left|v_{\omega k}^{\text{int}} \right|^2 \text{Re} \, G_{\omega k}^{\text{int}}  + v_{\omega k}^{\text{int}\,\dagger} \cdot
 \frac{i M_{\omega k}^2}{\det M_{\omega k}}\cdot v_{\omega k}^{\text{int}}\,\,\text{Im} \, G_{\omega k}^{\text{int}} \nonumber \\
  - \; & \; 4 \, \text{Re}
  \left[
  v_{\omega k}^{\text{ext}\,\dagger} \cdot \frac{i M_{\omega k}}{\det M_{\omega k}}\cdot
 v_{\omega k}^{\text{int}}\right] X_{\omega k}
  \, \text{Im} \, G_{\omega k}^{\text{ext}}  \,.
 \label{eq:osfinal}
\end{align}
To obtain this expression we have used the relation (\ref{eq:N2}) to express everything in terms of quantities in the right and future regions and we have used (\ref{eq:N1}) to relate quantities to their complex conjugates. We also simplified the final line in (\ref{eq:osfinal}) using the constraint (\ref{eq:rel}). Imposing this constraint explicitly causes the imaginary part of the action to vanish. Recall that the real and imaginary parts of the Green's function describe reactive and dissipative dynamics, respectively.

The holographic banner (\ref{eq:osfinal}) shows how the external and internal boundary data is correlated via the boundary Green's functions and the ratios in (\ref{eq:DX}). In classical or semiclassical regimes we can use this object to solve for the internal state in terms of external boundary sources, or vice versa. We will do so in the following subsection.

\subsection{Interior classical solution and semiclassical phase}
\label{sec:interior}

The on-shell action (\ref{eq:os4}) will obey the Hamilton-Jacobi equation with respect to both the future and past values of the fields, $\phi_{\omega k}^{(0)F}$ and $\phi_{\omega k}^{(0)P}$. This follows from the usual textbook arguments, where we are keeping the boundary values $\phi_{\omega k}^{(0)L}$ and $\phi_{\omega k}^{(0)R}$ fixed. That is, we have an initial-boundary value problem for the scalar field within the holographic banner.
The initial and final (that is, past and future) momenta are given by
\be\label{eq:pis}
\pi_{\omega k}^{(0)F} \equiv \frac{\pa S}{\pa \phi_{\omega k}^{(0)F}} \,, \quad \pi_{\omega k}^{(0)P} \equiv - \frac{\pa S}{\pa \phi_{\omega k}^{(0)P}} \,, \quad \bar \pi_{\omega k}^{(0)F} \equiv \frac{\pa S}{\pa \bar \phi_{\omega k}^{(0)F}} \,, \quad \bar \pi_{\omega k}^{(0)P} \equiv -\frac{\pa S}{\pa \bar \phi_{\omega k}^{(0)P}} \,.
\ee
Using (\ref{eq:pis}), the classical interior field is given  in terms of the initial and boundary data as
\begin{align}\label{eq:phiF}
\phi_{\omega k \, \text{cl}}^{(0)F} =
& \left(
\begin{array}{cc}
     e^{i \Theta_{\omega k}} &
     e^{-i \Theta_{\omega k}}
\end{array}
\right) \cdot \frac{v^\text{ext}_{\omega k}}{X_{\omega k}} \nonumber \\
& - \cos(2 \Theta_{\omega k}) \phi_{\omega k}^{(0)P}
  -  \frac{\sin(2 \Theta_{\omega k})}{\text{Im} \, G_{\omega k}^{\text{int}}} \left(\text{Re} \, G_{\omega k}^{\text{int}} \, \phi_{\omega k}^{(0)P} +  \bar \pi_{\omega k}^{(0)P} \right) \,.
\end{align}
Here we have also used (\ref{eq:rel}) to show that the interior classical state can be determined in terms of initial and boundary data without explicitly needing the
exterior Green's function, which is the basic object characterising response in the dual boundary field theory.
We have checked directly from the matching formulae in \S\ref{sec:hor} and \S\ref{sec:bdy}, with the momentum defined in the usual way from the Lagrangian and without using Hamilton-Jacobi theory, that (\ref{eq:phiF}) is the future classical solution. The Hamilton-Jacobi approach, however, will allow us easily to build semiclassical wavepackets.

In the remainder, to keep the expressions more transparent, we will restrict to the field in the classical `ground state' of the black hole, with no sources in the past.
This allows us to focus on how the holographic boundary determines the future interior state. That is we set
\be\label{eq:past}
\phi_{\omega k}^{(0)P} = \pi_{\omega k}^{(0)P} = 0 \,.
\ee
This is the (classical) thermofield double state of the dual boundary theory \cite{Maldacena:2001kr}. It will likely also be interesting, in the future, to consider states with excitations in the past, cf.~\cite{Balasubramanian:2022gmo}. For the case of no past excitations, (\ref{eq:phiF}) becomes
\be\label{eq:phiTFD}
\phi_{\omega k \, \text{cl}}^{(0)F} =
\left(
\begin{array}{cc}
     e^{i \Theta_{\omega k}} &
     e^{-i \Theta_{\omega k}}
\end{array}
\right) \cdot \frac{v^\text{ext}_{\omega k}}{X_{\omega k}}
 \,,
\ee
and the future momentum is similarly found to be
\be\label{eq:piTFD}
\pi_{\omega k \, \text{cl}}^{(0)F} =
\frac{v^{\text{ext}\, \dagger}_{\omega k}}{X_{\omega k}} \cdot \left(
\begin{array}{c}
     e^{-i \Theta_{\omega k}} \, {\overline{G^{\text{int}}_{\omega k
     }}}\\
     e^{i \Theta_{\omega k}} \, G^{\text{int}}_{\omega k
     }
\end{array}
\right) 
 \,.
\ee

The semiclassical limit of the interior quantum wavefunction is obtained by building a wavepacket that is strongly peaked on given past field values and momenta, such as (\ref{eq:past}). The boundary values $\phi_{\omega k}^{(0)L}$ and $\phi_{\omega k}^{(0)R}$ are kept fixed, and we will sometimes drop the explicit dependence on these.
We can quickly recall how the wavepacket works, only working to exponential order in the semiclassical limit. We can furthermore discuss each mode separately, with the full state given by
\be
\Psi\left[\phi_{\omega k}^{(0)F}\right] = \prod_{\omega k} \Psi_{\omega k}\left[\phi_{\omega k}^{(0)F}\right] \,.
\ee
To keep the expressions clean we will drop the $\w k$ subscripts in the following couple of formulae. A semiclassical Gaussian wavepacket is given by
\be\label{eq:gauss}
\Psi\left[\phi^{(0)F}\right] = \int d\phi^{(0)P} d\bar \phi^{(0)P} e^{-\frac{1}{\delta^2} |\phi^{(0) P}|^2} e^{i S[\phi^{(0)P}, \phi^{(0)F}]} \,.
\ee
Here $\delta$ is a small width about $\phi^{(0)P}=0$ and the absence of a phase proportional to $\phi^{(0)P}$ indicates that the wavepacket has vanishing past momentum.
The integral (\ref{eq:gauss}) is Gaussian and can therefore be performed exactly. Using our result (\ref{eq:osfinal}) for the holographic banner, the answer to exponential order is
\be\label{eq:psifinal}
\Psi\left[\phi^{(0)F}\right] = e^{- \chi \, \left|\phi^{(0)F} - \phi^{(0)F}_\text{cl}\right|^2} e^{ 2 i \text{Re} \left[ \pi^{(0)F}_\text{cl}\left(\phi^{(0)F} - \phi^{(0)F}_\text{cl} \right) \right]}e^{i S_\text{cl}} \,,
\ee
where $\{\phi^{(0)F}_\text{cl}, \pi^{(0)F}_\text{cl}\}$ is the classical solution in (\ref{eq:phiTFD}) and (\ref{eq:piTFD}) with a trivial past (\ref{eq:past}).
The state is strongly peaked on the classical solution.
The quantum mechanical features of (\ref{eq:psifinal}) are the spreading of the wavepacket, determined by $\chi$, and the phase $i S_\text{cl}$. We can now specify these.

The action (\ref{eq:osfinal}) on the classical solution gives, restoring the wavevector dependence,
\be\label{eq:scl}
S_{\omega k \, \text{cl}} =  \frac{v^{\text{ext}\, \dagger}_{\omega k}}{X_{\omega k}} \cdot
\left(
\begin{array}{cc}
\text{Re} \left[G_{\omega k}^{\text{int}} + X^2_{\omega k}  G_{\omega k}^{\text{ext}}\right] & e^{-2 i \Theta} \overline{G_{\omega k}^{\text{int}}}\\
 e^{2 i \Theta} G_{\omega k}^{\text{int}} & \text{Re} \left[G_{\omega k}^{\text{int}} + X^2_{\omega k}  G_{\omega k}^{\text{ext}}\right]
\end{array}
\right)
\cdot \frac{v^\text{ext}_{\omega k}}{X_{\omega k}} \,.
\ee
The spreading $\chi$ is given by
\be\label{eq:chi}
\chi_{\omega k} = \left(\frac{|\nu_{\omega k}|^2}{\frac{1}{\delta^2} - i s_{\omega k}''} - i s_{\omega k}''\right) \,.
\ee
The terms here come firstly from the variance of the action about the past (and future) data
\be\label{eq:spp}
s_{\omega k}'' \equiv \frac{\pa^2 s_{\omega k}}{\pa \phi_{\omega k}^{(0) P} \pa \bar \phi_{\omega k}^{(0) P}} = \frac{\pa^2 s_{\omega k}}{\pa \phi_{\omega k}^{(0) F} \pa \bar \phi_{\omega k}^{(0) F}} = \text{Re} \,G_{\omega k}^\text{int} + \cot(2 \Theta_{\omega k}) \, \text{Im} \, G_{\omega k}^\text{int} \,,
\ee
and secondly from a van Vleck type term evaluated on the classical solution
\be\label{eq:nu}
\nu_{\omega k} \equiv \frac{\pa \bar \pi_{\omega k}^{(0)P}}{\pa \phi_{\omega k}^{(0) F}} = - \frac{\text{Im}\, G^\text{int}_{\omega k}}{\sin (2 \Theta_{\omega k})}  \,,
\ee
Note that $\frac{\pa \bar \pi^{(0)P}}{\pa \phi^{(0) F}} = - \frac{\pa^2 s}{\pa \phi^{(0) F} \pa \bar \phi^{(0) P}} = - \frac{\pa \pi^{(0)F}}{\pa \bar \phi^{(0) P}}$. The above formulae for $s''_{\omega k}$ and $\nu_{\omega k}$ give a physical meaning to the interior Green's function as controlling wavefunction spreading.

It may be helpful to recap what we have just done. Using the holographic banner (\ref{eq:osfinal}) we have obtained the semiclassical state (\ref{eq:psifinal}) of the future interior. The classical state in the past interior (\ref{eq:past}) is trivial, so that the nontrivial classical state in the future interior, given by (\ref{eq:phiTFD}) and (\ref{eq:piTFD}), is due to the exterior boundary sources $v_{\omega k}^\text{ext}$ in (\ref{eq:vv}). These sources lead to the quantum mechanical phase (\ref{eq:scl}) of the future interior state. The spreading of the semiclassical wavepacket, in contrast, is determined by the background geometry and the width of the past wavepacket, according to (\ref{eq:chi}). In the following section we consider an explicit example of these phenomena.

\section{The near-singularity state}

\subsection{Holographic banner for a BTZ black hole}
\label{sec:oned}

We can evaluate the quantities in the holographic banner (\ref{eq:osfinal}) explicitly for the case of a BTZ black hole, which has $d=1$ in the metric (\ref{eq:met}). We noted above (\ref{eq:ns}) that the $k=0$ modes in $d=1$ have a logarithmic behaviour near the singularity, similar to higher dimensional cases. We will set $k=0$ in this discussion, and drop the $k$ label.

As is well-known, see e.g.~\cite{Birmingham:2001pj}, the solutions to the wave equation (\ref{eq:eom}) in a BTZ background are hypergeometric functions. We give details of the solution in Appendix \ref{app:btz}. From the formulae in the Appendix we find that to
leading order near the singularity, $z_\mathcal{H} \ll z_F$, the classical solution (\ref{eq:phiTFD}) and (\ref{eq:piTFD}) becomes 
\begin{align}\label{eq:FL}
\phi^{(0)F}_{\omega \, \text{cl}} &= p_\omega \log \frac{z_F}{z_{\mathcal H}}  + \cdots \,, \\
\pi^{(0)F}_{\omega \, \text{cl}} &= \bar p_\omega + \cdots \,,
\end{align}
where the `momentum' $p_\omega$ depends on the boundary sources as
\begin{align}\label{eq:PI}
p_\omega \equiv \frac{2 i}{\pi} \frac{z_\mathcal{H}^{1-\Delta}}{\Gamma(\Delta - 1)} & \left|\Gamma[\half(\Delta + i \omega z_\mathcal{H})]\right|^2 \\ \, & \times \, \left(\sinh[{\textstyle \frac{\pi}{2}}(\omega z_\mathcal{H} - i \Delta)] \phi^{(0) R}_\omega - \sinh[{\textstyle \frac{\pi}{2}}(\omega z_\mathcal{H} + i \Delta)] \phi^{(0) L}_\omega \right) \,. \nonumber
\end{align}
Recall that $\Delta$ is the scaling dimension of the AdS/CFT dual operator to the bulk field $\phi$.
Expanding (\ref{eq:PI}) at low frequencies gives $p_\w \propto T^{\Delta - 1} (\phi^{(0) R}_\omega+\phi^{(0) L}_\omega)$ while at high frequencies $p_\w \propto \omega^{\Delta - 1} (e^{-\frac{1}{2} i \pi \Delta}\phi^{(0) R}_\omega - e^{\frac{1}{2} i \pi \Delta} \phi^{(0) L}_\omega)$. Here $T$ is the temperature (\ref{eq:temp}). Given that the frequency-space boundary sources have CFT scaling dimension $[\phi_\omega^{(0) R/L}]= -\Delta$, these formulae suggest that the field in the future interior can be associated to a boundary CFT scaling dimension $[\phi_\omega^{(0) F}] = -1$.

To leading order near the singularity the classical action (\ref{eq:scl}) is found, using the formulae in Appendix \ref{app:btz},  to be 
\begin{align}\label{eq:SL}
S_\text{cl} & =  \left|p_\omega \right|^2 \log \frac{z_F}{z_{\mathcal H}} + \cdots \,,
\end{align}
and the spreading of the wavepacket in (\ref{eq:chi}) is determined by
\be\label{eq:snu}
s''_{\omega} = \frac{1}{\log \frac{z_F}{z_{\mathcal H}}} + \cdots \,, \qquad \text{and} \qquad
\nu_{\omega} = \frac{\pi}{2 \sin (\pi \Delta)} \frac{1}{\log^2 \frac{z_F}{z_{\mathcal H}}} + \cdots \,.
\ee
We noted previously that the spreading does not depend on the boundary sources. Here we see that for the BTZ black hole the spreading is also frequency-independent to leading order near the singularity.

It is instructive to introduce the interior clock
\be\label{eq:log}
\tau \equiv \log \frac{z_F}{z_\mathcal{H}} \,,
\ee
so that $\tau \to \infty$ near the singularity. The near-singularity interior state (\ref{eq:psifinal}) is then, setting $\phi^{(0) F} = Z$ to unencumber the notation,
\be\label{eq:psiinterior}
\Psi\left[Z,\tau\right] = e^{- \chi(\tau) \, \left|Z - p \tau\right|^2} e^{i \left( \bar p Z + p \bar Z\right)}e^{- i |p|^2 \tau} \,,
\ee
with the spreading of the norm $\left|\Psi[Z,\tau]\right|^2$ controlled by
\be\label{eq:spread}
\text{Re}\, \chi(\tau) = \frac{\pi^2}{4 \sin^2(\pi \Delta)} \frac{\delta^2}{\tau^2} \frac{1}{\delta^4 + \tau^2}   \,. 
\ee
The state (\ref{eq:psiinterior}) is almost that of a conventional free quantum mechanical particle in two dimensional flat space. Indeed, we can see back in the action 
(\ref{eq:actwk}) that the $|\phi_\omega|^2$ `mass terms' are subleading for the field modes near the singularity, leading to the free particle behaviour. However, the late time $\tau^4$ growth of the variance in (\ref{eq:spread}) is faster than the conventional $\tau^2$ growth for a free particle following the Schr\"odinger equation.

The origin of the fast spreading is the `van Vleck' term (\ref{eq:nu}) which is a kind of past/future susceptibility. This term relates the spreading in the future to the spread of the wavepacket in the past. If different cutoffs are used near the future and past singularities, as explained in footnote \ref{foot:unequal}, one finds that in the BTZ van Vleck term (\ref{eq:snu}) the logarithm squared splits up as $\log^2 \frac{z_F}{z_\mathcal{H}} \to \log \frac{z_F}{z_\mathcal{H}} \times \log \frac{z_P}{z_\mathcal{H}}$ in the limit $z_F, z_P \gg z_\mathcal{H}$.
Thus if the past cutoff $z_P$ is kept fixed as $z_F \to \infty$, then the expected $\tau^2$ growth of the variance of the future wavepacket is recovered. Said differently, the fast variance growth in (\ref{eq:spread}) is an artifact of simultaneously evolving the initial and final slices.

\subsection{Higher dimensions and the exterior-interior map}

Near a higher-dimensional Schwarzschild singularity, the leading behaviour of the modes is again logarithmic in the $z$ coordinate, as we saw in (\ref{eq:ns}). Similarly to the BTZ case we have just considered, the mass term drops out of the action (\ref{eq:actwk}) near the singularity, resulting in free particle motion in the time variable $\tau$ defined in (\ref{eq:log}). The near-singularity behaviour we have just described for the BTZ black hole thus carries over to higher dimensions, where it now holds for all $\omega$ and $k$. In particular, the near-singularity wavefunction resulting from a Gaussian wavepacket in the past has the same form as (\ref{eq:psiinterior}),
\be
\Psi[Z,\tau] = \prod_{\omega k} e^{- \chi_{\omega k}(\tau) \, \left|Z_{\omega k} - p_{\omega k} \tau\right|^2} e^{i \left( \bar p_{\omega k} Z_{\omega k} + p_{\omega k} \bar Z_{\omega k}\right)}e^{- i |p_{\omega k}|^2 \tau} \,.
\ee
The `momenta' $p_{\omega k}$ will again be linear functions of the boundary data $\left\{\phi^{(0)L}_{\omega k}, \phi^{(0) R}_{\omega k} \right\}$. The explicit linear relation depends on the solution to the bulk wave equation and will be more complicated than the BTZ expression (\ref{eq:PI}). The spreading $\chi_{\omega k}$ is still given by (\ref{eq:chi}), (\ref{eq:spp}) and (\ref{eq:nu}). It is independent of the boundary data but depends on the form of the interior Green's functions for the higher dimensional backgrounds.

With a given state in the past, the holographic banner therefore defines a map
\be\label{eq:map1}
\left\{\phi^{(0)L}_{\omega k}, \phi^{(0) R}_{\omega k} \right\} \qquad \mapsto \qquad |\Psi(\tau)\rangle \in \bigotimes_{\omega k} {\mathcal H}_{\omega k} \,,
\ee
from classical boundary data to interior near-singularity single-particle states living in the free particle Hilbert spaces ${\mathcal H}_{\omega k}$. It should be emphasised that this is an interior bulk Hilbert space and that the `time' $\tau$ under which the interior state evolves is not the boundary time but a relational bulk time (see e.g.~\cite{Hartnoll:2022snh}). The entire black hole interior is described on the boundary by a single time-independent state, as we explained around Fig.~\ref{fig:bigpicture} in the introduction. It is not known, at present, what boundary relational observable would capture the evolution in interior bulk time $\tau$. See e.g.~\cite{Araujo-Regado:2022gvw, deBoer:2022zps, Leutheusser:2025zvp} for ideas in related directions.

The free particle Hilbert spaces in (\ref{eq:map1}) are just those of the non-interacting modes of the scalar field in the bulk. In the following section we will recall the emergence of single-particle dynamics near the singularity in a far more nontrivial setting.

\section{Interior BKL chaos and mixing time}
\label{sec:bkl}

The BKL scenario describes gravitational dynamics close to a spacelike singularity \cite{Belinsky:1970ew, Damour:2002et, belinski_henneaux_2017}. Key classical and semiclassical aspects of this dynamics have recently been summarised in \cite{DeClerck:2023fax, Hartnoll:2025hly, DeClerck:2026vgd}. The basic degree of freedom is the $(d+1)$-metric $\gamma_{ij}(x)$ on spatial slices. Near to the singularity each point $x$ on the spatial slice decouples and evolves independently. Semiclassical states thus have the form
\be\label{eq:xprod}
\Psi[\gamma_{ij}] \approx \prod_x \psi_x(\gamma_{ij}(x)) \,.
\ee
The $\half (d+1)(d+2)$ independent components of $\gamma_{ij}$ can be split up into (i) a relational `time' $\tau$ that diverges towards the singularity, this variable controls the collapse of the local volume element $\sqrt{\gamma}$, (ii) $d$ scale factors that we will label $\vec z$ and (iii) $\half d(d+1)$ remaining `off-diagonal' components. For a suitable parametrisation it is found that the off-diagonal modes freeze towards the singularity while the scale factors $\vec z$ evolve with respect to $\tau$ following a $\tau$-independent hyperbolic billiards Hamiltonian.

For concreteness, we focus on pure gravity in a four dimensional bulk, so that $d=2$. To introduce familiar coordinates, set $\vec z = (x,y)$. In this case the hyperbolic billiard is half of the fundamental domain of $SL(2,\Z)$ in ${\mathbb H}_2$, i.e. the particle is constrained to
\be\label{eq:domain}
0 < x < \half \quad \text{and} \quad x^2 + y^2 \geq 1 \,,
\ee
and the dynamics within the domain is given by
\be\label{eq:maass}
\pi_\tau^2 = y^2 \left(\pi_{x}^2 + \pi_{y}^2 \right) \,.
\ee
This billiard problem is the canonical example of arithmetic chaos --- an ergodic single-particle dynamics with additional arithmetic structure due to conserved Hecke operators \cite{Bogomolny:1992cj}.

\subsection{The interior mixing time}

The Hamilton-Jacobi equation corresponding to (\ref{eq:maass}) is
\be
y^2 \left[\left(\frac{\pa S}{\pa x}\right)^2 + \left(\frac{\pa S}{\pa y}\right)^2 \right] = \left(\frac{\pa S}{\pa \tau}\right)^2 \,.
\ee
This equation is solved using separation of variables as
\be\label{eq:acts}
S = - \vep \tau + k x \pm \vep \left[\sqrt{1 - \frac{k^2 y^2}{\vep^2}} - \text{arctanh} \left(\sqrt{1 - \frac{k^2 y^2}{\vep^2}} \right) \right] \,.
\ee
Here $\vep$ and $k$ are constants of integration giving the conserved `energy' and `momentum' of the motion on ${\mathbb H}_2$. The term in square brackets in (\ref{eq:acts}) is just the WKB limit of the Bessel functions $K_{i\vep}(ky)$ appearing in the full solution to the Laplace equation on the billiard domain in e.g.~\cite{Hartnoll:2025hly}. Ignoring the domain boundaries (\ref{eq:domain}) to start with, the classical solutions are then obtained from
\be\label{eq:xk}
x_{o} = \frac{\pa S}{\pa k} \,, 
\ee
where $x_{o}$ is constant. This gives the familiar circular hyperbolic geodesics,
\be\label{eq:xcl}
x = x_{\text{cl}} \equiv x_{o} \mp \sqrt{\frac{\vep^2}{k^2} - y^2}
\,.
\ee

A semiclassical wavepacket supported on the classical trajectory (\ref{eq:xcl}) with energy $\vep$ and $k=k_o$ is built as
\begin{align}
\psi_{\vep, k_o}(x,y,\tau) & = \int dk \, e^{-(k - k_o)^2/[2 (\Delta k)^2] - i (k - k_o) x_o + i S} \label{eq:wav1} \\
& = e^{i S_\text{cl} + i k_o (x - x_\text{cl}) - \chi_\text{cl} (x - x_\text{cl})^2} \,. 
\end{align}
Here $S_\text{cl}$ is the action (\ref{eq:acts}) evaluated on the classical solution (\ref{eq:xcl}) with $k=k_o$. The integral has been performed by expanding the exponent to quadratic order in $k-k_o$ and then doing the Gaussian integral.
The spreading in $x$ is determined by the real part
\be\label{eq:Dx}
(\Delta x)^2 =  \frac{1}{2 \, \text{Re} \, \chi_\text{cl}} = \frac{1}{(\Delta k)^2} + \frac{\vep^4 (\Delta k)^2}{k_o^6 \left(\frac{\vep^2}{k_o^2} - y^2\right)} \,.
\ee
The first term in (\ref{eq:Dx}) is the spread built into the wavepacket in (\ref{eq:wav1}), while the second term is the dynamical spreading. We focus on this second term in the remainder and will thereby recover the well-known exponential divergence of geodesics in hyperbolic space.

The proper transverse spread about the geodesic is
\be
(\Delta w)^2 = \frac{(\Delta x)^2}{y^2} \frac{(x - x_o)^2}{(\vep/k_o)^2} = \frac{\vep^2 (\Delta k)^2}{k_o^4} \frac{1}{y^2} \,.
\ee
In the first equality the factor of $1/y^2$ comes from the billiard metric
$ds^2 = (dx^2 + dy^2)/y^2$ on the upper half plane. This hyperbolic metric underlies \eqref{eq:maass}. The factor of
$\frac{(x - x_o)}{(\vep/k_o)} \Delta x$ is the projection of the $\Delta x$ separation onto the direction perpendicular to the geodesic circle (\ref{eq:xcl}), as $\frac{(x - x_o)}{(\vep/k_o)}$ is the cosine of the angle parametrising the circle. The second equality uses the second term in (\ref{eq:Dx}) and the equation (\ref{eq:xcl}) for the circle. Integrating the proper distance $ds = \sqrt{x'(y)^2+1} \frac{dy}{y}$ along the geodesic, measured from the apex at $x=x_o$, gives the relation $y = \frac{\vep}{k_o} \text{sech}\, s$ and hence
\be\label{eq:growth}
\Delta w = \frac{\Delta k}{k_o} \cosh s \,.
\ee
Thus we recover the standard result that the transverse spread of the quantum wavepacket grows exponentially in the proper distance $s$ traveled. When the geodesic reaches a domain boundary (\ref{eq:domain}) it must be reflected back
into the fundamental domain. At the three boundaries the reflections are given respectively, in terms of $z \equiv x + i y$, by $z \mapsto - \bar z$, $z \mapsto 1 - \bar z$ and $z \mapsto 1/\bar z$. These reflections 
(or `bounces') are isometries and do not interrupt the exponential growth (\ref{eq:growth}) of the wavepacket, as was illustrated in Fig.~\ref{fig:domain} in the introduction.

We can express the growth (\ref{eq:growth}) in terms of the relational time $\tau$ along a classical trajectory using the remaining equation of motion
\be\label{eq:ept}
- \t_o = \frac{\pa S}{\pa \vep} \,. 
\ee
This leads to
\be
y = y_\text{cl} \equiv \frac{\vep}{k_o} \sech(\tau - \tau_o) \,, \qquad \Rightarrow \qquad s = \tau - \tau_o \,.
\label{eq:st}
\ee
In fact it was already clear from (\ref{eq:maass}) that $\tau$ is the proper hyperbolic length. The Hamilton-Jacobi equation of motion (\ref{eq:ept}) can be obtained from the quantum mechanical state together with (\ref{eq:xk}) above by extending (\ref{eq:wav1}) to be a wavepacket over both $k$ and $\vep$. Integrating over $k$ and $\vep$ then produces a wavepacket supported on both (\ref{eq:xcl}) and (\ref{eq:st}).\footnote{It is worth noting that under $(\vep,k) \to \lambda (\vep,k)$ the action (\ref{eq:acts}) scales as $S \to \lambda S$. This means that there is no quadratic change to the action under a shift by $(\delta \vep,\delta k) \propto (\vep,k)$ and hence no dynamical spreading in that direction. To obtain a conventional Gaussian wavepacket with (\ref{eq:ept}) and (\ref{eq:xk}) as saddle point equations it is essential to include explicit widths $\Delta k$ and/or $\Delta \vep$ in the wavepacket.}

The billiard (\ref{eq:domain}) has a finite, order one proper area. It follows from (\ref{eq:growth}) and (\ref{eq:st}) that by the mixing (or, Ehrenfest) time
\be\label{eq:mix}
\tau_\text{mix} \equiv \log \frac{k_o}{\Delta k} \,,
\ee
the wavepacket will have spread out across the entire billiard. Beyond this time the wavefunction rapidly becomes uniform throughout the domain, so that memory of the initial state is lost. The uniform spread is a consequence of `quantum unique ergodicity' \cite{marklof2004arithmetic}. We should emphasise that the mixing (\ref{eq:mix}) involves the relational interior time $\tau$ and is unrelated to scrambling that occurs as a function of boundary time \cite{Hartman:2013qma, Shenker:2013pqa, Susskind:2014rva, Stanford:2014jda, Brown:2015lvg}. BKL mixing occurs much deeper in the black hole interior.

If the interior can be probed using a relational boundary observable, ergodic mixing over the timescale (\ref{eq:mix}) is a fundamental effect to uncover. The chaotic evolution towards the singularity ends at $\tau_\text{mix}$ in an equilibrium quantum state of the bulk scale factors $\vec z$. This final state is highly quantum in the sense of having a large variance, but can also be semiclassical if the curvature of spacetime has not reached the Planck or string scale. This scenario was mentioned in \cite{DeClerck:2023fax}, but we now have the tools to be more precise. To estimate the curvature at $\tau_\text{mix}$ we can relate $\tau$ to the local volume $\sqrt{\gamma}$ of the spatial slice. The explicit Milne-like change of variables given in \cite{Hartnoll:2025hly} expresses
\be
\log \sqrt{\gamma} = - e^{\tau} \frac{1 - x + x^2 + y^2}{\sqrt{2} y} \,.
\ee
The local volume can then be evaluated along the classical geodesics using the equations of motion (\ref{eq:xcl}) and (\ref{eq:st}) to obtain $x(\tau)$ and $y(\tau)$. It follows that at late times
\be\label{eq:rootgamma}
\sqrt{\gamma} \approx c_1 e^{- c_2 \, e^{2 \tau}} \qquad \Rightarrow \qquad \left. \sqrt{\gamma} \right|_\text{mix} \sim c_1 e^{- c_2 \left(\frac{k_o}{\Delta k}\right)^2} \,.
\ee
Here $c_1, c_2$ are constants that depend on the trajectory. The volume collapses doubly exponentially with the relational time $\tau$ and hence the volume at which the universe is quantum mechanically smeared is exponentially small in $\frac{k_o}{\Delta k}$. This needs to be compared to the Planck length $\ell_\text{Pl}$, as we now discuss.

Given a state $|\text{past}\rangle$ in the past, such as the thermofield double state, the future interior state (\ref{eq:xprod}) at each point $x$ will be a function of the boundary sources. Thus the holographic banner now defines a map
\be\label{eq:map2}
\left\{\phi^{(0)L}, \phi^{(0) R}, |\text{past}\rangle \right\} \qquad \mapsto \qquad |\Psi(\tau)\rangle \in \bigotimes_{x} {\mathcal H}_{x} \,,
\ee
At any given spatial point $x$, a semiclassical universe is a wavepacket of the form (\ref{eq:wav1}). 
The map (\ref{eq:map2}) is far more complicated than our previous (\ref{eq:map1}) for a scalar field, as the
classical evolution from past and boundary to the future interior is now nonlinear. Furthermore, as we have just seen, the late time interior evolution by $\tau$ is no longer free but quantum chaotic.

Something that we did learn by considering the simpler case of a scalar field, however, is how the constants of integration and the variance of the future interior wavepacket are determined. The energy and momenta $\vep, k_o$ of the classical trajectory guiding the wavepacket (\ref{eq:wav1}) will depend on the boundary sources. These boundary sources, however, are fixed and do not introduce the quantum variance $\Delta k$. The quantum variance instead originates in the initial state $|\text{past}\rangle$. In semiclassical initial states such as the thermofield double, quantum fluctuations are suppressed relative to the classical expectation values by a factor of $G_N \sim \ell_\text{Pl}^2$. Thus a very crude estimate is
\be\label{eq:ratio}
\frac{\left. \sqrt{\gamma} \right|_\text{mix}}{\ell_\text{Pl}^3} \sim \frac{\ell_\text{BH}^3}{\ell_\text{Pl}^3} e^{- \ell^2_\text{BH}/\ell^2_\text{Pl}} \ll 1\,.
\ee
Here $\ell_\text{BH}$ is a scale in the classical geometry, which could be the AdS scale or a scale associated to the horizon radius. Eq.~(\ref{eq:ratio}) says that the universe hits the Planck scale well before the quantum variance dominates. Even though the mixing is exponentially fast, the collapse of the volume is doubly exponentially fast and therefore wins. Thus quantum gravity and stringy corrections will become important before wavefunction spreading.

To turn the logic around, we could ask how much initial variance is necessary to have quantum mixing before the interior universe collapses to the Planck scale. From (\ref{eq:rootgamma}) we see that this requires
\be\label{eq:bound}
\frac{(\Delta k)^2}{k_o^2} \; \gtrsim \; \frac{1}{\log (\ell_\text{BH}/\ell_\text{Pl})} \,.
\ee
A variance of this magnitude is still compatible with a classical exterior with $1 \gg \frac{\Delta k}{k_o}$. Classical initial states obeying (\ref{eq:bound}) lead to highly quantum, but semiclassical, black hole interiors. The smearing of the trajectories across the domain does not resolve the singularity, as beyond the mixing time the volume continues to collapse and the trace of the extrinsic curvature continues to grow. However, a mixing time that is accessible within the semiclassical bulk regime presents itself as a natural target for boundary probes of the black hole interior. It may therefore be interesting to construct holographic states with the larger variance required by (\ref{eq:bound}).

It can also be noted that the classical mixing time is much faster then the quantum mixing time we have just discussed. The Lyapunov exponent from (\ref{eq:growth}) and (\ref{eq:st}) is $\lambda_L = 1$. This rate of divergence of classical trajectories is uniform throughout the billiard and hence the classical mixing time is just $\tau_\text{mix\ cl} \equiv 1/\lambda_L = 1$. This time is easily reached before the interior universe becomes Planck-sized and may also have an interesting dual field-theoretic manifestation. In particular, beyond this classical mixing time, time-averaged properties of the classical interior should become largely insensitive to the choice of boundary and initial conditions.

\subsection{Ensembles of boundary theories}
\label{sec:ensemble}

A different way to obtain a large variance is to allow
the boundary theory to be part of an ensemble. This ensemble is not a fundamental aspect of the duality, as in \cite{Saad:2019lba}, but rather a tool within a conventional holographic dictionary in which we average over the boundary data $\{\phi^{(0)L}, \phi^{(0)R}\}$. In particular, the ensemble can involve a distribution of boundary couplings that is wide enough to require a classical averaging over geometries in the bulk. The ensemble of boundary couplings will result in a classical average over the $\vep, k_o$ quantum numbers of the wavepackets (\ref{eq:wav1}). Let the distribution of these couplings be $P(\vep,k_o)$ and have `classical' widths $\Delta_\text{cl} \vep$ and $\Delta_\text{cl} k_o \gg \Delta k$.

All interior expectation values are to be averaged as
\be
\overline{\langle \ocal \rangle} \equiv \int d\vep dk_o  P(\vep,k_o) \langle \ocal \rangle_{\vep k_o} \,.
\ee
The full variance of the momentum $\pi_x$ is then
\be\label{eq:vartot}
\overline{\langle \pi_x^2 \rangle} - \overline{\langle \pi_x \rangle}^2 = \overline{\phantom{\langle}\text{var} \, \pi_x \phantom{\rangle}} + \text{var}_\text{cl} \langle \pi_x \rangle = (\Delta k)^2 + (\Delta_\text{cl} k_o)^2 \approx (\Delta_\text{cl} k_o)^2 \,.
\ee
The second expression here is a sum of the classical average of the quantum variance and the classical variance of the quantum average. All that (\ref{eq:vartot}) is saying is that quantities must be calculated over a range of classical trajectories controlled by the ensemble width. These trajectories will spread over the entire billiard domain at the ensemble mixing time
\be\label{eq:ensmix}
\tau_\text{ens mix} \equiv \log \frac{\overline{k_o}}{\Delta_\text{cl} k_o} \,.
\ee
This time can be a much earlier time than (\ref{eq:mix}), so that mixing can now occur before the local volume reaches the Planck scale.

\section{Discussion}
\label{sec:dis}

Holographic banners distill the connection between boundary sources and interior bulk states into a single physical quantity. The Hamilton-Jacobi formulation we have employed to construct the interior states is just the semiclassical limit of the gravitational path integral. A fully quantum mechanical holographic banner can be defined as
\be\label{eq:z}
Z[\phi^{(0)L},\phi^{(0)R},\phi^{(0)F},\phi^{(0)P}] \equiv \int_{\curvedsquare} {\mathcal D} \phi \; e^{i S[\phi]} \,.
\ee
The path integral is over all fields $\phi$ inside the banner with boundary values $\phi^{(0)}$. The integral is, of course, over the entire field content of the bulk theory, including the metric and any excited stringy modes. The quantum holographic banner (\ref{eq:z}) combines the basic object in the AdS/CFT dictionary \cite{Witten:1998qj, Gubser:1998bc} with the Hartle-Hawking path integral construction of the wavefunction of the universe \cite{Hartle:1983ai}. Within the full path integral the topology of the bulk slices is allowed to fluctuate. The construction can also be extended to allow the boundary to have more (or fewer) than two disconnected components.

A further generalisation is to define the past state on a slice at some finite boundary time, rather than in the infinite past. This precludes the need to consider the `white hole' region of the full spacetime. This would also be a natural setting to use a Euclidean preparation of the initial state, in the spirit of \cite{Hartle:1983ai, Maldacena:2001kr}.

Our detailed discussion of a scalar field in a fixed background involved a convenient choice of coordinates, attuned to the symmetry of the background. However, it is important that the holographic banner is defined only by the boundary values of the fields. It does not require any particular choice of radial coordinate or causal structure in the bulk. Indeed, the gravitational backreaction of sources can change the causality properties of the holographic banner leading to long \cite{Shenker:2013pqa, Shenker:2013yza} or traversable \cite{Gao:2016bin} wormholes. These effects do not alter the definition of the holographic banner.

In the holographic banner the left/right and future/past data appear on an equal footing. This fact is entirely emergent from the perspective of the boundary theory and we do not currently posses a relational boundary definition of the interior quantum states, which evolve in an emergent time and in an emergent Hilbert space. In quantum cosmology, relational observables are usually anchored in the Hamiltonian constraint. This is a gauge constraint implementing invariance under bulk diffeomorphisms and therefore does not play nicely with the manifestly gauge-invariant holographic dictionary. Holographic boundary theories, on the other hand, often have an $SU(N)$ gauge symmetry. In simple models, such as the fuzzy sphere, the matrix $SU(N)$ symmetry is related to a geometrical diffeomorphism symmetry in the large $N$ limit. Relational matrix observables in that simple context have been explored recently in \cite{Fliss:2025kzi}. One place to look for interior observables, then, may be within the matrix degrees of freedom of the boundary theory. Intriguingly, the arithmetic chaos arising in the interior BKL dynamics does have deep links to random matrices \cite{rudnick} and it would fascinating if these could be connected to the matrix degrees of freedom in a holographic dual.

Holographic banners may also be useful beyond an AdS/CFT setting. In particular, in a de Sitter universe an object like (\ref{eq:z}) bridges the divide between the static patch and the meta-observer \cite{Anninos:2012qw}. The future and past values of the fields define the Hartle-Hawking state \cite{Hartle:1983ai} while the left and right values now define sources along a static patch worldline, as considered in \cite{Anninos:2011af}. The holographic banner captures how worldline sources change the future Hartle-Hawking state.

If the interior states have their own holographic description --- e.g.~via a putative dS/CFT \cite{Strominger:2001pn, Maldacena:2002vr} or primon gas \cite{Hartnoll:2025hly, DeClerck:2025mem} correspondence --- then one can also run the relational question in the opposite direction. What are the relational observables in the holographic dual to the wavefunction that re-discover the static patch or the asymptotically AdS boundary?

\section*{Acknowledgements}

It is a pleasure to thank Matt Headrick for a helpful remark about averaging ensembles, and Marine De Clerck for discussions at the start of this project. We also thank Julian Sonner and Balt Van Rees for helpful discussions about holographic duals to the Schwinger-Keldysh contour. This work has been partially supported by STFC consolidated grant ST/T000694/1. SAH is partially supported by Simons Investigator award \#620869 and MJB is supported by a Gates Cambridge Scholarship (OPP1144). 

\appendix

\section{Example of matching across horizons}
\label{app:match}

Consider the globally regular mode $e^{i \kappa U}$, with $\kappa>0$. In the near-horizon regions this can be expanded in the modes (\ref{eq:phiUV}) as
\be\label{eq:fourier2}
e^{i \kappa U} = \int_{-\infty}^\infty \frac{d\omega}{2 \pi} B(\omega) e^{i \omega \log |U|} \,.
\ee
Inverting the Fourier transform gives
\be
B(\omega) = e^{\text{sgn}(U) \frac{\pi \omega}{2}} \kappa^{i \omega} \Gamma(-i \omega) \,.
\ee
Here the $\text{sgn}(U)$ tells us which region we are in.
The frequency integral in (\ref{eq:fourier2}) should be taken just above the real axis. We see that indeed $B(\omega)$ matches within $P/R$ and $F/L$, as stated in (\ref{eq:A2}), but differs between regions where $U$ has a different sign.

\section{Details of the BTZ modes}
\label{app:btz}

The solutions to the wave equation (\ref{eq:eom}) when $d=1$ are given by hypergeometric functions. We will set $k=0$. The ingoing modes, as normalised in (\ref{eq:full}), are 
\begin{align}
\phi^{R-}_\omega(z) = & a_{\omega} \hat z^\Delta \left(1 - \hat z^2 \right)^{\frac{i \hat \omega}{2}} {}_2F_1 \left(\frac{\Delta + i \hat \omega}{2}, \frac{\Delta + i \hat \omega}{2}, \Delta, \hat z^2 \right) \nn \\
& + b_{\omega} \hat z^{2-\Delta} \left(1 - \hat z^2 \right)^{\frac{i \hat \omega}{2}} {}_2F_1\left(\frac{2-\Delta + i \hat \omega }{2}, \frac{2-\Delta + i \hat \omega}{2}, 2-\Delta, \hat z^2 \right) \,, \label{eq:btzR}
\end{align}
here we set $\hat z = z/z_{\mathcal H}$ and $\hat \omega = \omega z_{\mathcal H}$.
The mass is related to the scaling dimension $\Delta$ of the dual operator by $(L m)^2 = \Delta(\Delta-2)$. The constants
\be
a_{\omega} = \left(\frac{2}{z_{\mathcal H}} \right)^{\frac{i \hat \omega}{2}}\frac{\Gamma(1-\Delta) \Gamma(1 - i \hat \omega)}{\Gamma(\half [2 - \Delta - i \hat \omega])^2} \,, \quad b_{\omega} =  \left(\frac{2}{z_{\mathcal H}} \right)^{\frac{i \hat \omega}{2}} \frac{\Gamma(\Delta-1) \Gamma(1 - i \hat \omega)}{\Gamma(\half [\Delta - i \hat \omega])^2} \,.
\ee

Expanding (\ref{eq:btzR}) near $z=0$ and comparing to (\ref{eq:nb}) gives the well-known expression for the retarded Green's function \cite{Son:2002sd}
\be\label{eq:btzext1}
G^{R-}_\omega = z_{\mathcal H}^{2-2 \Delta}\frac{2 \sin (\pi \Delta)}{\pi} \left(\frac{\Gamma(2-\Delta) \Gamma(\half [\Delta - i \hat \omega])}{\Gamma(\half [2-\Delta - i \hat \omega])}\right)^2 \,.
\ee
Here and in the following we freely use the Gamma function reflection formula. The normalisation factor is found to be
\be\label{eq:btzext2}
N^{R-}_\omega =z_{\mathcal H}^{\Delta-2} \sqrt{L}  \left(\frac{2}{z_{\mathcal H}} \right)^{\frac{i \hat \omega}{2}}  \frac{\Gamma(\Delta - 1) \Gamma(1 - i \hat \omega)}{\Gamma(\half [\Delta - i \hat \omega])^2} \,.
\ee
These expressions are verified to obey (\ref{eq:jr}).

The modes in the interior that we need are instead
\begin{align}
\phi^{F+}_\omega(z) = & a_{-\omega} \hat z^\Delta \left(\hat z^2 - 1\right)^{\frac{i \hat \omega}{2}} {}_2F_1 \left(\frac{\Delta + i \hat \omega}{2}, \frac{\Delta + i \hat \omega}{2}, \Delta, \hat z^2 \right) \nn \\
& + b_{-\omega} \hat z^{2-\Delta} \left(\hat z^2 - 1 \right)^{\frac{i \hat \omega}{2}} {}_2F_1\left(\frac{2-\Delta + i \hat \omega }{2}, \frac{2-\Delta + i \hat \omega}{2}, 2-\Delta, \hat z^2 \right) \,. \label{eq:btzF}
\end{align}
The constants are the same as in (\ref{eq:btzR}) but with $\omega \to - \omega$.
Expanding (\ref{eq:btzF}) at large $z$ we may now read off the interior Green's function and normalisation from (\ref{eq:ns}). One must deal carefully with the logarithmic running. The answer is most compactly given as
\be\label{eq:i1}
N^{F+}_\omega = \sqrt{L}  \left(\frac{2}{z_{\mathcal H}} \right)^{1-\frac{i \hat \omega}{2}} \frac{\Gamma(1 + i \hat \omega)}{\Gamma(\half[2 - \Delta + i \hat \omega]) \Gamma(\half [\Delta + i \hat \omega])} \frac{1}{G^{F+}_\omega} \,,
\ee
and
\begin{align}
\frac{1}{G^{F+}_\omega}=  \log \frac{z_F}{z_{\mathcal H}} - \gamma_E - \text{Re} \, \psi(\half[\Delta + i \hat \omega]) - \frac{i \pi}{2} \coth \left[{\textstyle \frac{\pi}{2}} (\hat \omega + i \Delta)\right]   \,. \label{eq:i2}
\end{align}
Here the digamma function $\psi(z) = \Gamma'(z)/\Gamma(z)$ and $\gamma_E$ is the Euler-Mascheroni constant. We used the reflection formula for the digamma function. One can verify that (\ref{eq:i1}) and (\ref{eq:i2}) together imply that (\ref{eq:jf}) is obeyed.

From (\ref{eq:btzext1}), (\ref{eq:btzext2}), (\ref{eq:i1}) and (\ref{eq:i2}) we can extract all the information needed to evaluate the holographic banner. Using the definitions given in (\ref{eq:DX}),
\be\label{eq:AA}
e^{i \Theta_{\omega}} = \frac{-i \sinh \left[{\textstyle \frac{\pi}{2}} (\hat \omega + i \Delta)\right]}{|\sinh \left[{\textstyle \frac{\pi}{2}} (\hat \omega + i \Delta)\right]|} + \ocal\left(\frac{1}{\log \frac{z_F}{z_{\mathcal H}}} \right) \,,
\ee
and
\be\label{eq:BB}
X_\omega = \frac{z_{\mathcal H}^{\Delta-1}}{2} \frac{\pi  \Gamma(\Delta - 1)}{|\sinh(\frac{\pi}{2}(\hat \omega + i \Delta))\, \Gamma[\frac{1}{2}(\Delta + i \hat \omega)]^2|} \frac{1}{\log \frac{z_F}{z_{\mathcal H}}} + \ocal\left(\frac{1}{\log^2 \frac{z_F}{z_{\mathcal H}}} \right)  \,.
\ee
Here we have expanded to leading order in the near-singularity limit $z_F \gg z_{\mathcal H}$. Similarly expanding the interior Green's function we obtain
\begin{align}\label{eq:CC}
\text{Re} \, G^{F+}_\omega & = \frac{1}{\log \frac{z_F}{z_{\mathcal H}}} + \left(\gamma_E + \text{Re} \, \psi(\half[\Delta + i \hat \omega]) + \frac{\pi}{4} \frac{\sin(\pi \Delta)}{|\sinh\left[{\textstyle \frac{\pi}{2}} (\hat \omega + i \Delta)\right]|^2} \right) \frac{1}{\log^2 \frac{z_F}{z_{\mathcal H}}} \cdots \,, \\ \text{Im} \, G^{F+}_\omega & = \frac{\pi}{4} \frac{\sinh(\pi \hat \omega)}{|\sinh\left[{\textstyle \frac{\pi}{2}} (\hat \omega + i \Delta)\right]|^2} \frac{1}{\log^2 \frac{z_F}{z_{\mathcal H}}} + \cdots \,.
\end{align}
From the expressions (\ref{eq:AA}), (\ref{eq:BB}) and (\ref{eq:CC}) it is straightforward to obtain the near-singularity results stated in \S\ref{sec:oned}.
In fact, from the formulae given it is also straightforward to compute the finite correction to the growing phase $S_\text{cl}$ in (\ref{eq:SL}) in the $z_F \gg z_{\mathcal H}$ limit, but the expression is messy and does not appear to be illuminating.

\providecommand{\href}[2]{#2}\begingroup\raggedright\endgroup


\begin{thebibliography}{10}

\bibitem{Fidkowski:2003nf}
L.~Fidkowski, V.~Hubeny, M.~Kleban and S.~Shenker, {{The Black hole singularity
  in AdS / CFT}}, \href{http://dx.doi.org/10.1088/1126-6708/2004/02/014}{JHEP
  {\bf 02}, 014, 2004},
  [\href{http://arxiv.org/abs/arXiv:hep-th/0306170}{{arXiv:hep-th/0306170}}].

\bibitem{Festuccia:2005pi}
G.~Festuccia and H.~Liu, {{Excursions beyond the horizon: Black hole
  singularities in Yang-Mills theories. I.}},
  \href{http://dx.doi.org/10.1088/1126-6708/2006/04/044}{JHEP {\bf 04}, 044,
  2006},
  [\href{http://arxiv.org/abs/arXiv:hep-th/0506202}{{arXiv:hep-th/0506202}}].

\bibitem{Frenkel:2020ysx}
A.~Frenkel, S.~A. Hartnoll, J.~Kruthoff and Z.~D. Shi, {{Holographic flows from
  CFT to the Kasner universe}},
  \href{http://dx.doi.org/10.1007/JHEP08(2020)003}{JHEP {\bf 08}, 003, 2020},
  [\href{http://arxiv.org/abs/arXiv:2004.01192}{{arXiv:2004.01192 [hep-th]}}].

\bibitem{Grinberg:2020fdj}
M.~Grinberg and J.~Maldacena, {{Proper time to the black hole singularity from
  thermal one-point functions}},
  \href{http://dx.doi.org/10.1007/JHEP03(2021)131}{JHEP {\bf 03}, 131, 2021},
  [\href{http://arxiv.org/abs/arXiv:2011.01004}{{arXiv:2011.01004 [hep-th]}}].

\bibitem{Afkhami-Jeddi:2025wra}
N.~Afkhami-Jeddi, S.~Caron-Huot, J.~Chakravarty and A.~Maloney, {{Imprint of
  the black hole singularity on thermal two-point functions}},  2025,
  [\href{http://arxiv.org/abs/arXiv:2510.21673}{{arXiv:2510.21673 [hep-th]}}].

\bibitem{Dodelson:2025jff}
M.~Dodelson, C.~Iossa and R.~Karlsson, {{Bouncing off a stringy singularity}},
  2025, [\href{http://arxiv.org/abs/arXiv:2511.09616}{{arXiv:2511.09616
  [hep-th]}}].

\bibitem{deBoer:2022zps}
J.~de~Boer, D.~L. Jafferis and L.~Lamprou, {{On black hole interior
  reconstruction, singularities and the emergence of time}},  2022,
  [\href{http://arxiv.org/abs/arXiv:2211.16512}{{arXiv:2211.16512 [hep-th]}}].

\bibitem{Leutheusser:2025zvp}
S.~Leutheusser and H.~Liu, {{Volume as an index of a subalgebra}},  2025,
  [\href{http://arxiv.org/abs/arXiv:2508.00056}{{arXiv:2508.00056 [hep-th]}}].

\bibitem{Hartnoll:2022snh}
S.~A. Hartnoll, {{Wheeler-DeWitt states of the AdS-Schwarzschild interior}},
  \href{http://dx.doi.org/10.1007/JHEP01(2023)066}{JHEP {\bf 01}, 066, 2023},
  [\href{http://arxiv.org/abs/arXiv:2208.04348}{{arXiv:2208.04348 [hep-th]}}].

\bibitem{DeWitt:1967yk}
B.~S. DeWitt, {{Quantum Theory of Gravity. 1. The Canonical Theory}},
  \href{http://dx.doi.org/10.1103/PhysRev.160.1113}{Phys. Rev. {\bf 160},
  1113--1148, 1967}.

\bibitem{deBoer:1999tgo}
J.~de~Boer, E.~P. Verlinde and H.~L. Verlinde, {{On the holographic
  renormalization group}},
  \href{http://dx.doi.org/10.1088/1126-6708/2000/08/003}{JHEP {\bf 08}, 003,
  2000},
  [\href{http://arxiv.org/abs/arXiv:hep-th/9912012}{{arXiv:hep-th/9912012}}].

\bibitem{deHaro:2000vlm}
S.~de~Haro, S.~N. Solodukhin and K.~Skenderis, {{Holographic reconstruction of
  space-time and renormalization in the AdS / CFT correspondence}},
  \href{http://dx.doi.org/10.1007/s002200100381}{Commun. Math. Phys. {\bf 217},
  595--622, 2001},
  [\href{http://arxiv.org/abs/arXiv:hep-th/0002230}{{arXiv:hep-th/0002230}}].

\bibitem{Heemskerk:2010hk}
I.~Heemskerk and J.~Polchinski, {{Holographic and Wilsonian Renormalization
  Groups}}, \href{http://dx.doi.org/10.1007/JHEP06(2011)031}{JHEP {\bf 06},
  031, 2011}, [\href{http://arxiv.org/abs/arXiv:1010.1264}{{arXiv:1010.1264
  [hep-th]}}].

\bibitem{Faulkner:2010jy}
T.~Faulkner, H.~Liu and M.~Rangamani, {{Integrating out geometry: Holographic
  Wilsonian RG and the membrane paradigm}},
  \href{http://dx.doi.org/10.1007/JHEP08(2011)051}{JHEP {\bf 08}, 051, 2011},
  [\href{http://arxiv.org/abs/arXiv:1010.4036}{{arXiv:1010.4036 [hep-th]}}].

\bibitem{McGough:2016lol}
L.~McGough, M.~Mezei and H.~Verlinde, {{Moving the CFT into the bulk with $
  T\overline{T} $}}, \href{http://dx.doi.org/10.1007/JHEP04(2018)010}{JHEP {\bf
  04}, 010, 2018},
  [\href{http://arxiv.org/abs/arXiv:1611.03470}{{arXiv:1611.03470 [hep-th]}}].

\bibitem{Hartman:2018tkw}
T.~Hartman, J.~Kruthoff, E.~Shaghoulian and A.~Tajdini, {{Holography at finite
  cutoff with a $T^2$ deformation}},
  \href{http://dx.doi.org/10.1007/JHEP03(2019)004}{JHEP {\bf 03}, 004, 2019},
  [\href{http://arxiv.org/abs/arXiv:1807.11401}{{arXiv:1807.11401 [hep-th]}}].

\bibitem{Blacker:2023ezy}
M.~J. Blacker and S.~Ning, {{Wheeler DeWitt states of a charged AdS$_{4}$ black
  hole}}, \href{http://dx.doi.org/10.1007/JHEP12(2023)002}{JHEP {\bf 12}, 002,
  2023}, [\href{http://arxiv.org/abs/arXiv:2308.00040}{{arXiv:2308.00040
  [hep-th]}}].

\bibitem{Blacker:2023oan}
M.~J. Blacker and S.~A. Hartnoll, {{Cosmological quantum states of de
  Sitter-Schwarzschild are static patch partition functions}},
  \href{http://dx.doi.org/10.1007/JHEP12(2023)025}{JHEP {\bf 12}, 025, 2023},
  [\href{http://arxiv.org/abs/arXiv:2304.06865}{{arXiv:2304.06865 [hep-th]}}].

\bibitem{Skenderis:2008dh}
K.~Skenderis and B.~C. van Rees, {{Real-time gauge/gravity duality}},
  \href{http://dx.doi.org/10.1103/PhysRevLett.101.081601}{Phys. Rev. Lett. {\bf
  101}, 081601, 2008},
  [\href{http://arxiv.org/abs/arXiv:0805.0150}{{arXiv:0805.0150 [hep-th]}}].

\bibitem{Skenderis:2008dg}
K.~Skenderis and B.~C. van Rees, {{Real-time gauge/gravity duality:
  Prescription, Renormalization and Examples}},
  \href{http://dx.doi.org/10.1088/1126-6708/2009/05/085}{JHEP {\bf 05}, 085,
  2009}, [\href{http://arxiv.org/abs/arXiv:0812.2909}{{arXiv:0812.2909
  [hep-th]}}].

\bibitem{Balasubramanian:1998sn}
V.~Balasubramanian, P.~Kraus and A.~E. Lawrence, {{Bulk versus boundary
  dynamics in anti-de Sitter space-time}},
  \href{http://dx.doi.org/10.1103/PhysRevD.59.046003}{Phys. Rev. D {\bf 59},
  046003, 1999},
  [\href{http://arxiv.org/abs/arXiv:hep-th/9805171}{{arXiv:hep-th/9805171}}].

\bibitem{Maldacena:2001kr}
J.~M. Maldacena, {{Eternal black holes in anti-de Sitter}},
  \href{http://dx.doi.org/10.1088/1126-6708/2003/04/021}{JHEP {\bf 04}, 021,
  2003},
  [\href{http://arxiv.org/abs/arXiv:hep-th/0106112}{{arXiv:hep-th/0106112}}].

\bibitem{Balasubramanian:2022gmo}
V.~Balasubramanian, A.~Lawrence, J.~M. Magan and M.~Sasieta, {{Microscopic
  Origin of the Entropy of Black Holes in General Relativity}},
  \href{http://dx.doi.org/10.1103/PhysRevX.14.011024}{Phys. Rev. X {\bf 14},
  011024, 2024},
  [\href{http://arxiv.org/abs/arXiv:2212.02447}{{arXiv:2212.02447 [hep-th]}}].

\bibitem{Hartnoll:2016apf}
S.~A. Hartnoll, A.~Lucas and S.~Sachdev, {{Holographic quantum matter}},  2016,
  [\href{http://arxiv.org/abs/arXiv:1612.07324}{{arXiv:1612.07324 [hep-th]}}].

\bibitem{Belinsky:1970ew}
V.~A. Belinsky, I.~M. Khalatnikov and E.~M. Lifshitz, {{Oscillatory approach to
  a singular point in the relativistic cosmology}},
  \href{http://dx.doi.org/10.1080/00018737000101171}{Adv. Phys. {\bf 19},
  525--573, 1970}.

\bibitem{Damour:2002et}
T.~Damour, M.~Henneaux and H.~Nicolai, {{Cosmological billiards}},
  \href{http://dx.doi.org/10.1088/0264-9381/20/9/201}{Class. Quant. Grav. {\bf
  20}, R145--R200, 2003},
  [\href{http://arxiv.org/abs/arXiv:hep-th/0212256}{{arXiv:hep-th/0212256}}].

\bibitem{belinski_henneaux_2017}
V.~Belinski and M.~Henneaux, \emph{The Cosmological Singularity}.
\newblock Cambridge Monographs on Mathematical Physics. CUP, 2017,
  \href{http://dx.doi.org/10.1017/9781107239333}{10.1017/9781107239333}.

\bibitem{Hartnoll:2025hly}
S.~A. Hartnoll and M.~Yang, {{The conformal primon gas at the end of time}},
  \href{http://dx.doi.org/10.1007/JHEP07(2025)281}{JHEP {\bf 07}, 281, 2025},
  [\href{http://arxiv.org/abs/arXiv:2502.02661}{{arXiv:2502.02661 [hep-th]}}].

\bibitem{DeClerck:2025mem}
M.~De~Clerck, S.~A. Hartnoll and M.~Yang, {{Wheeler-DeWitt wavefunctions for 5d
  BKL dynamics, automorphic L-functions and complex primon gases}},
  \href{http://dx.doi.org/10.1007/JHEP11(2025)160}{JHEP {\bf 11}, 160, 2025},
  [\href{http://arxiv.org/abs/arXiv:2507.08788}{{arXiv:2507.08788 [hep-th]}}].

\bibitem{Belin:2020oib}
A.~Belin, A.~Lewkowycz and G.~Sarosi, {{Gravitational path integral from the
  $T^2$ deformation}}, \href{http://dx.doi.org/10.1007/JHEP09(2020)156}{JHEP
  {\bf 09}, 156, 2020},
  [\href{http://arxiv.org/abs/arXiv:2006.01835}{{arXiv:2006.01835 [hep-th]}}].

\bibitem{Araujo-Regado:2022gvw}
G.~Araujo-Regado, R.~Khan and A.~C. Wall, {{Cauchy slice holography: a new
  AdS/CFT dictionary}}, \href{http://dx.doi.org/10.1007/JHEP03(2023)026}{JHEP
  {\bf 03}, 026, 2023},
  [\href{http://arxiv.org/abs/arXiv:2204.00591}{{arXiv:2204.00591 [hep-th]}}].

\bibitem{Unruh:1976db}
W.~G. Unruh, {{Notes on black hole evaporation}},
  \href{http://dx.doi.org/10.1103/PhysRevD.14.870}{Phys. Rev. D {\bf 14}, 870,
  1976}.

\bibitem{Herzog:2002pc}
C.~P. Herzog and D.~T. Son, {{Schwinger-Keldysh propagators from AdS/CFT
  correspondence}}, \href{http://dx.doi.org/10.1088/1126-6708/2003/03/046}{JHEP
  {\bf 03}, 046, 2003},
  [\href{http://arxiv.org/abs/arXiv:hep-th/0212072}{{arXiv:hep-th/0212072}}].

\bibitem{Liu:2018crr}
H.~Liu and J.~Sonner, {{Holographic systems far from equilibrium: a review}},
  \href{http://dx.doi.org/10.1088/1361-6633/ab4f91}{Rept. Prog. Phys. {\bf 83},
  016001, 2019},
  [\href{http://arxiv.org/abs/arXiv:1810.02367}{{arXiv:1810.02367 [hep-th]}}].

\bibitem{Brown:2015lvg}
A.~R. Brown, D.~A. Roberts, L.~Susskind, B.~Swingle and Y.~Zhao, {{Complexity,
  action, and black holes}},
  \href{http://dx.doi.org/10.1103/PhysRevD.93.086006}{Phys. Rev. D {\bf 93},
  086006, 2016},
  [\href{http://arxiv.org/abs/arXiv:1512.04993}{{arXiv:1512.04993 [hep-th]}}].

\bibitem{Gubser:1998bc}
S.~S. Gubser, I.~R. Klebanov and A.~M. Polyakov, {{Gauge theory correlators
  from noncritical string theory}},
  \href{http://dx.doi.org/10.1016/S0370-2693(98)00377-3}{Phys. Lett. B {\bf
  428}, 105--114, 1998},
  [\href{http://arxiv.org/abs/arXiv:hep-th/9802109}{{arXiv:hep-th/9802109}}].

\bibitem{Witten:1998qj}
E.~Witten, {{Anti de Sitter space and holography}},
  \href{http://dx.doi.org/10.4310/ATMP.1998.v2.n2.a2}{Adv. Theor. Math. Phys.
  {\bf 2}, 253--291, 1998},
  [\href{http://arxiv.org/abs/arXiv:hep-th/9802150}{{arXiv:hep-th/9802150}}].

\bibitem{Son:2002sd}
D.~T. Son and A.~O. Starinets, {{Minkowski space correlators in AdS / CFT
  correspondence: Recipe and applications}},
  \href{http://dx.doi.org/10.1088/1126-6708/2002/09/042}{JHEP {\bf 09}, 042,
  2002},
  [\href{http://arxiv.org/abs/arXiv:hep-th/0205051}{{arXiv:hep-th/0205051}}].

\bibitem{Doroshkevich:1978aq}
A.~G. Doroshkevich and I.~D. Novikov, {{Space-Time and Physical Fields inside a
  Black Hole}}, {Zh. Eksp. Teor. Fiz. {\bf 74}, 3--12, 1978}.

\bibitem{Fournodavlos:2018lrk}
G.~Fournodavlos and J.~Sbierski, {{Generic Blow-Up Results for the Wave
  Equation in the Interior of a Schwarzschild Black Hole}},
  \href{http://dx.doi.org/10.1007/s00205-019-01434-0}{Arch. Ration. Mech. Anal.
  {\bf 235}, 927--971, 2020},
  [\href{http://arxiv.org/abs/arXiv:1804.01941}{{arXiv:1804.01941 [gr-qc]}}].

\bibitem{Birmingham:2001pj}
D.~Birmingham, I.~Sachs and S.~N. Solodukhin, {{Conformal field theory
  interpretation of black hole quasinormal modes}},
  \href{http://dx.doi.org/10.1103/PhysRevLett.88.151301}{Phys. Rev. Lett. {\bf
  88}, 151301, 2002},
  [\href{http://arxiv.org/abs/arXiv:hep-th/0112055}{{arXiv:hep-th/0112055}}].

\bibitem{DeClerck:2023fax}
M.~De~Clerck, S.~A. Hartnoll and J.~E. Santos, {{Mixmaster chaos in an AdS
  black hole interior}}, \href{http://dx.doi.org/10.1007/JHEP07(2024)202}{JHEP
  {\bf 07}, 202, 2024},
  [\href{http://arxiv.org/abs/arXiv:2312.11622}{{arXiv:2312.11622 [hep-th]}}].

\bibitem{DeClerck:2026vgd}
M.~De~Clerck, {{An introduction to BKL theory}},
  \href{http://dx.doi.org/10.22323/1.498.0002}{PoS {\bf Modave2024}, 002,
  2026}.

\bibitem{Bogomolny:1992cj}
E.~B. Bogomolny, B.~Georgeot, M.~J. Giannoni and C.~Schmit, {{Arithmetical
  chaos}}, \href{http://dx.doi.org/10.1016/S0370-1573(97)00016-1}{Phys. Rept.
  {\bf 291}, 219--324, 1997}.

\bibitem{marklof2004arithmetic}
J.~Marklof, {Arithmetic quantum chaos},  in \emph{Encyclopedia of Mathematical
  Physics, Vol 1}, pp.~212--220.
\newblock Amsterdam: Elsevier, 2004.

\bibitem{Hartman:2013qma}
T.~Hartman and J.~Maldacena, {{Time Evolution of Entanglement Entropy from
  Black Hole Interiors}}, \href{http://dx.doi.org/10.1007/JHEP05(2013)014}{JHEP
  {\bf 05}, 014, 2013},
  [\href{http://arxiv.org/abs/arXiv:1303.1080}{{arXiv:1303.1080 [hep-th]}}].

\bibitem{Shenker:2013pqa}
S.~H. Shenker and D.~Stanford, {{Black holes and the butterfly effect}},
  \href{http://dx.doi.org/10.1007/JHEP03(2014)067}{JHEP {\bf 03}, 067, 2014},
  [\href{http://arxiv.org/abs/arXiv:1306.0622}{{arXiv:1306.0622 [hep-th]}}].

\bibitem{Susskind:2014rva}
L.~Susskind, {{Computational Complexity and Black Hole Horizons}},
  \href{http://dx.doi.org/10.1002/prop.201500092}{Fortsch. Phys. {\bf 64},
  24--43, 2016}, [\href{http://arxiv.org/abs/arXiv:1403.5695}{{arXiv:1403.5695
  [hep-th]}}].

\bibitem{Stanford:2014jda}
D.~Stanford and L.~Susskind, {{Complexity and Shock Wave Geometries}},
  \href{http://dx.doi.org/10.1103/PhysRevD.90.126007}{Phys. Rev. D {\bf 90},
  126007, 2014}, [\href{http://arxiv.org/abs/arXiv:1406.2678}{{arXiv:1406.2678
  [hep-th]}}].

\bibitem{Saad:2019lba}
P.~Saad, S.~H. Shenker and D.~Stanford, {{JT gravity as a matrix integral}},
  2019, [\href{http://arxiv.org/abs/arXiv:1903.11115}{{arXiv:1903.11115
  [hep-th]}}].

\bibitem{Hartle:1983ai}
J.~B. Hartle and S.~W. Hawking, {{Wave Function of the Universe}},
  \href{http://dx.doi.org/10.1103/PhysRevD.28.2960}{Phys. Rev. D {\bf 28},
  2960--2975, 1983}.

\bibitem{Shenker:2013yza}
S.~H. Shenker and D.~Stanford, {{Multiple Shocks}},
  \href{http://dx.doi.org/10.1007/JHEP12(2014)046}{JHEP {\bf 12}, 046, 2014},
  [\href{http://arxiv.org/abs/arXiv:1312.3296}{{arXiv:1312.3296 [hep-th]}}].

\bibitem{Gao:2016bin}
P.~Gao, D.~L. Jafferis and A.~C. Wall, {{Traversable Wormholes via a Double
  Trace Deformation}}, \href{http://dx.doi.org/10.1007/JHEP12(2017)151}{JHEP
  {\bf 12}, 151, 2017},
  [\href{http://arxiv.org/abs/arXiv:1608.05687}{{arXiv:1608.05687 [hep-th]}}].

\bibitem{Fliss:2025kzi}
J.~R. Fliss, A.~Frenkel, S.~A. Hartnoll and R.~M. Soni, {{Minimal areas from
  entangled matrices}},
  \href{http://dx.doi.org/10.21468/SciPostPhys.18.6.171}{SciPost Phys. {\bf
  18}, 171, 2025},
  [\href{http://arxiv.org/abs/arXiv:2408.05274}{{arXiv:2408.05274 [hep-th]}}].

\bibitem{rudnick}
Z.~Rudnick, {Zeta functions in arithmetic and their spectral statistics}, {
  Lectures at DMV Summer School,
  \url{http://www.math.tau.ac.il/~rudnick/papers/ihp.ps}, 2000}.

\bibitem{Anninos:2012qw}
D.~Anninos, {{De Sitter Musings}},
  \href{http://dx.doi.org/10.1142/S0217751X1230013X}{Int. J. Mod. Phys. A {\bf
  27}, 1230013, 2012},
  [\href{http://arxiv.org/abs/arXiv:1205.3855}{{arXiv:1205.3855 [hep-th]}}].

\bibitem{Anninos:2011af}
D.~Anninos, S.~A. Hartnoll and D.~M. Hofman, {{Static Patch Solipsism:
  Conformal Symmetry of the de Sitter Worldline}},
  \href{http://dx.doi.org/10.1088/0264-9381/29/7/075002}{Class. Quant. Grav.
  {\bf 29}, 075002, 2012},
  [\href{http://arxiv.org/abs/arXiv:1109.4942}{{arXiv:1109.4942 [hep-th]}}].

\bibitem{Strominger:2001pn}
A.~Strominger, {{The dS / CFT correspondence}},
  \href{http://dx.doi.org/10.1088/1126-6708/2001/10/034}{JHEP {\bf 10}, 034,
  2001},
  [\href{http://arxiv.org/abs/arXiv:hep-th/0106113}{{arXiv:hep-th/0106113}}].

\bibitem{Maldacena:2002vr}
J.~M. Maldacena, {{Non-Gaussian features of primordial fluctuations in single
  field inflationary models}},
  \href{http://dx.doi.org/10.1088/1126-6708/2003/05/013}{JHEP {\bf 05}, 013,
  2003},
  [\href{http://arxiv.org/abs/arXiv:astro-ph/0210603}{{arXiv:astro-ph/0210603}}].

\end{thebibliography}
\end{document}